\documentclass[runningheads]{llncs}
\usepackage{eccv}
\usepackage{eccvabbrv}
\usepackage{graphicx}
\usepackage{booktabs}
\usepackage{multirow}
\usepackage{amsmath}
\usepackage{mathtools}
\usepackage[accsupp]{axessibility}
\usepackage{longtable}
\usepackage[table]{xcolor}
\usepackage{csquotes}
\usepackage[breaklinks,colorlinks,citecolor=eccvblue]{hyperref}
\usepackage[pass]{geometry}

\DeclareMathOperator{\sgn}{sgn}

\begin{document}

\title{NURBS Splatting: A Unified Differentiable Rendering Framework for Vector Graphics}
\titlerunning{NURBS Splatting}
\author{Jingye Qiu \and Shizhe Zhou\thanks{Corresponding author.}}
\authorrunning{J. Qiu and S. Zhou}
\institute{Hunan University, Changsha, China\\\email{anicoder@outlook.com, shizhe@hnu.edu.cn}}
\maketitle

\begin{abstract}
  Differentiable rendering of planar rational splines remains largely underexplored, despite their widespread use in vector graphics and design. Existing differentiable vector renderers primarily focus on B\'{e}zier curves and rely on analytic rasterization, which can suffer from gradient instability and limited flexibility. We propose NURBS Splatting, a unified framework that represents planar rational curves as continuous Gaussian fields. By sampling Gaussians along the curve parameter domain and inside closed regions, rendering is reformulated as a smooth accumulation process with stable gradients. Our method naturally supports long splines, rational weights, non-uniform knots, and closed-region filling. We demonstrate its effectiveness in calligraphy reconstruction, vectorization frameworks, and long-spline image abstraction, showing improved stability and reconstruction quality over existing approaches.
\end{abstract}
\section{Introduction}
Non-Uniform Rational B-Splines (NURBS)~\cite{10.5555/265261} are the standard curve and surface representation in CAD and manufacturing.
Every major CAD kernel---SolidWorks, CATIA, Rhino, Siemens NX---and exchange formats such as STEP (ISO~10303) and IGES use NURBS, and isogeometric analysis~\cite{hughesIsogeometricAnalysisCAD2005} builds directly on them for finite-element simulation.
Their advantage over polynomial B-splines is well established: rational weights and non-uniform knots let NURBS exactly represent conic sections---circles, ellipses, parabolic arcs---that polynomial splines can only approximate~\cite{10.5555/265261,farinCurvesSurfacesCAGD2007}, while also providing per-control-point shape modulation for tighter local geometric control.

As pixel-based image generation models improve, differentiable rendering offers a promising way to convert these pixel outputs into physical, manufacturable vector formats.
Because splines are compact and resolution-independent, they are highly suited for such tasks. Berio \etal~\cite{berioNeuralImageAbstraction2025}, for example, showed that optimizing long smoothing splines can produce high-quality artistic abstractions that can be reproduced with robotic pen plotters.
However, spline representations remain only partially explored in differentiable vector graphics. Current differentiable 2D renderers~\cite{liDifferentiableVectorGraphics2020,liu2025bzier} are restricted to B'ezier curves, while Berio \etal~\cite{berioNeuralImageAbstraction2025} support B-splines but do not introduce rational weights or non-uniform knots.
Consequently, differentiable renderers cannot faithfully represent shapes that CAD systems natively support---such as exact circular arcs and conic profiles---and remain disconnected from the NURBS-based modeling infrastructure that industry has relied on for decades.

We address this with \textbf{NURBS Splatting}.
We adaptively sample isotropic Gaussians along NURBS curves---positions, widths, and analytical derivatives computed via the Cox--de Boor recursion---and inside closed regions via an SDF-modulated grid, then composite them through a tile-based Gaussian splatting rasterizer.
The pipeline is fully differentiable: image-space losses drive gradients back to control points, rational weights, knot intervals, stroke widths, colors, and opacities.
For filled closed curves, grid-step annealing progressively refines spatial resolution during optimization.

The main contributions of this work are as follows:
\begin{itemize}
  \item The first differentiable renderer for NURBS curves in 2D image space, supporting joint optimization of rational weights and non-uniform knot vectors.
  \item A Gaussian splatting formulation for both open stroke contours and filled closed regions, combining adaptive arc-length sampling with SDF-modulated interior fill and grid-step annealing.
  \item Evaluation on calligraphy reconstruction, image vectorization, and neural image abstraction showing that rational splines improve both quality and speed over polynomial baselines.
\end{itemize}

\begin{figure}[tb]
  \centering
  \includegraphics[width=1.0\linewidth]{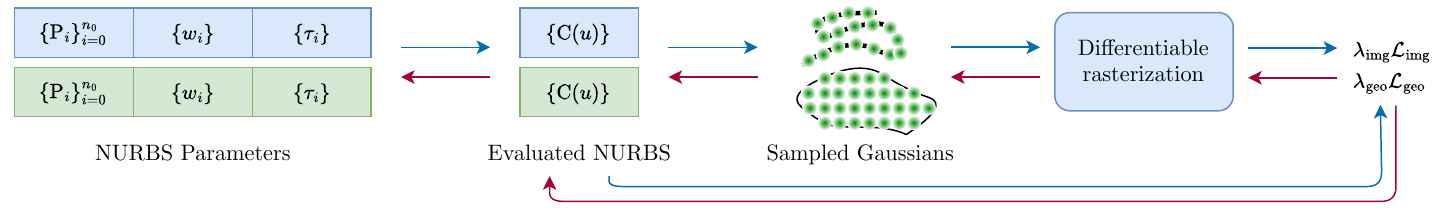}
  \caption{We optimize NURBS parameters (control points $\mathbf{P}_i$, weights $w_i$, knot intervals $\tau_i$) by sampling isotropic Gaussians along the curve and interior. These primitives are rendered via a differentiable tile-based rasterizer, enabling end-to-end optimization through image and geometric losses.}
  \label{fig:pipeline}
\end{figure}

\section{Related Work}
\subsection{Differentiable Rendering}
DiffVG \cite{liDifferentiableVectorGraphics2020} pioneered differentiable rasterization, which enabled gradient-based optimization of vector images.
A variety of applications have since been explored, including image vectorization \cite{maLayerwiseImageVectorization2022, chakrabortyImageVectorizationGradient2025,hirschornOptimizeReduceTopDown2024,wangLayeredImageVectorization2025}, text-guided drawing synthesis \cite{fransCLIPDrawExploringTexttoDrawing2022,vinkerCLIPassoSemanticallyAwareObject2022}, text-to-SVG generation via score distillation \cite{jainVectorFusionTexttoSVGAbstracting2023, xingDiffSketcherTextGuided2023, xingSVGDreamerTextGuided2024}, semantic typography \cite{iluzWordAsImageSemanticTypography2023}, collage generation \cite{wangImageSpaceCollagePacking2025}, animation generation \cite{zhangSketchDancingTextDrivenFramework2025}, sketch refinement \cite{tianSketchRefinerTextGuidedSketch2025}, \etc.
However, the DiffVG approach relies on per-pixel intersection tests, making high-resolution optimization slow and memory-intensive.

An alternative line of work uses Gaussian primitives for rendering. In 3D vision, Kerbl \etal~\cite{kerbl3DGaussianSplatting2023} propose 3D Gaussian Splatting for novel-view synthesis. B\'{e}zierGS~\cite{maBezierGSDynamicUrban2025} and SplineGS~\cite{yoonSplineGSLearningSmooth2024} further extend this paradigm with spline-based motion modeling.
The idea has since been adapted to 2D domains. Huang \etal~\cite{huang2DGaussianSplatting2024} introduce 2D Gaussian Splatting for radiance-field reconstruction, collapsing the 3D volume into oriented 2D Gaussian disks to enforce view-consistent geometry. Likewise, Zhang \etal~\cite{zhangGaussianImage1000FPS2025} present GaussianImage, encoding images directly as collections of 2D Gaussians to achieve ultra-fast rendering and compression. In the context of vector graphics, Liu \etal~\cite{liu2025bzier} propose B\'{e}zier Splatting, which samples 2D Gaussian kernels along B\'ezier curves and rasterizes via a Gaussian-splatting pipeline. These works demonstrate that Gaussian splatting is a powerful differentiable primitive for both 3D scene reconstruction and 2D vector graphics rendering.

Notably, all the aforementioned methods employ either polynomial B\'{e}zier curves or polynomial B-splines as the underlying geometric primitives. To the best of our knowledge, no existing differentiable rendering framework supports rational spline primitives for 2D image-space optimization.

\subsection{Non-Uniform Rational B-Splines}
NURBS \cite{10.5555/265261} generalize polynomial B-splines by introducing rational weights and non-uniform knot vectors, allowing exact representation of conic sections, analytic curves, and flexible free-form shapes. They have become the de facto industry standard for modeling curves and surfaces in CAD and graphics. While classical algorithms like the Cox--de Boor recursion \cite{coxNumericalEvaluationBSplines1972,deBoorCalculatingBSplines1972} evaluate and sample points for rendering, many modern professional tools employ GPU curve rasterization based on the implicit form of rational B\'ezier curves obtained from NURBS via knot insertion \cite{loopResolutionIndependentCurve2005}. Beyond geometric modeling, NURBS have found broader scientific impact through isogeometric analysis \cite{hughesIsogeometricAnalysisCAD2005}, which unifies CAD geometry and finite-element analysis by using NURBS basis functions directly as the computational basis.

Recent work has begun to integrate NURBS into differentiable and learning-based pipelines.
Deva Prasad \etal \cite{devaprasadNURBSDiffDifferentiableProgramming2022} introduce NURBS-Diff, a differentiable NURBS layer that provides forward and backward passes through the NURBS evaluation process, enabling tasks such as curve and surface fitting, surface offsetting, \etc.
Worchel and Alexa~\cite{worchelDifferentiableRenderingParametric2023} propose differentiable rendering of parametric geometry, including B-spline surfaces, by directly deriving screen-space gradients for parametric primitives.
Tojo \etal~\cite{tojoFabricable3DWire2024} optimize B-spline curves via differentiable rendering for fabricable 3D wire art.
On the generative side, SplineGen~\cite{zouSplineGenApproximatingUnorganized2025} leverages a generative model to fit B-spline surfaces to unorganized point clouds, NeuroNURBS~\cite{fanNeuroNURBSLearningEfficient2024} learns compact NURBS surface representations for 3D solids, and BrepGen~\cite{xuBrepGenBrepGenerative2024} employs diffusion models to generate B-rep CAD models whose faces are NURBS surfaces.
Despite this growing interest, existing differentiable and generative NURBS methods predominantly target 3D geometry fitting or CAD reconstruction rather than image-based rendering.
\section{Method}
Our pipeline is illustrated in \cref{fig:pipeline}.
We first define NURBS curve primitives with learnable control points, rational weights, and knot vectors (\cref{sec:nurbs_primitives}).
Isotropic Gaussians are then adaptively sampled along curve contours (\cref{sec:gaussian_contour_sampling}) or within closed filled regions via an SDF-modulated grid (\cref{sec:interior_filling}).
These Gaussians are composited through a tile-based splatting rasterizer to produce a differentiable image (\cref{sec:diff_splatting}), and image-space losses drive gradients back to all curve parameters (\cref{sec:optimization}).
Although we focus on NURBS, our Gaussian sampling formulation is generic and readily extends to other parametric primitives.

\subsection{NURBS Primitives}
\label{sec:nurbs_primitives}
Each curve in our system is a NURBS curve of degree $p$ (order $k{=}p{+}1$) with $n+1$ control points:
\begin{equation}
\mathbf{C}(u) = \frac{\sum_{i=0}^{n} N_{i,p}(u)\, w_i \, \mathbf{P}_i}{\sum_{i=0}^{n} N_{i,p}(u)\, w_i}=\frac{\mathbf{S}(u)}{W(u)}, 
\quad u \in [u_0, u_m],
\end{equation}
where $\{\mathbf{P}_i \in \mathbb{R}^2\}$ are control points, $\{w_i \in \mathbb{R}^+\}$ are rational weights, and $\{N_{i,p}(u)\}$ are B-spline basis functions defined over a non-decreasing knot vector
$\mathbf{U} = \{u_0, u_1, \dots, u_m\}$.
The basis functions $N_{i,p}(u)$ are defined by the Cox--de Boor recursion~\cite{coxNumericalEvaluationBSplines1972,deBoorCalculatingBSplines1972}, which guarantees compact support and $C^{p-1}$ continuity.

The curve derivative, needed for arc-length estimation and regularization, follows by the quotient rule:
\begin{equation}
\label{eq:curve_derivative}
\mathbf{C}'(u)
= \frac{
W(u) \sum_{i=0}^{n} N'_{i,p}(u)\, w_i \, \mathbf{P}_i
- \mathbf{S}(u) \sum_{i=0}^{n} N'_{i,p}(u)\, w_i
}{W^2(u)},
\end{equation}
where the basis-function derivatives $N'_{i,p}$ are also obtainable via the Cox--de Boor recursion.
The arc length used in \cref{sec:gaussian_contour_sampling} can be approximated by numerical quadrature:
\begin{equation}
\label{eq:arc_length}
L = \int_{u_p}^{u_{m-p}} \|\mathbf{C}'(u)\|_2 \, du \approx \sum_{j=1}^{N_L} \|\mathbf{C}'(u_j)\|_2 \, \mathrm{\Delta} u.
\end{equation}

Rather than optimizing control points and knots directly, we leverage key points $\{\mathbf{K}_i\}_{i=1}^{n_k}$ as the primary learnable parameters and derive control points and knot vectors from them.

For open curves, the control points coincide with the key points, and the knot vector is clamped at both endpoints with multiplicity $k$.
Only the $n_k - k + 1$ interior knot intervals $\{\tau_j > 0\}$ are stored as learnable parameters; the full knot vector is reconstructed by prepending and appending $p$ zero-intervals (to enforce endpoint multiplicity) and taking a cumulative sum, which inherently guarantees monotonicity.

For closed curves, the control polygon is formed by cyclically wrapping the last $\lceil p/2 \rceil$ and first $\lfloor p/2 \rfloor$ key points to produce $n_k + p$ control points.
We store $n_k$ core knot intervals; the periodic knot vector is then built by wrapping the last $p$ core intervals before and the first $p$ after, followed by a cumulative sum.

Each curve also carries an RGB color $\mathbf{c} \in [0,1]^3$ and an opacity $o \in [0,1]$.
We extend the control points to $\mathbb{R}^3$ by appending a third coordinate encoding local stroke width, so evaluating $\mathbf{C}(u)$ simultaneously yields position and width at parameter~$u$.

\subsection{Gaussian Sampling on Contours}
\label{sec:gaussian_contour_sampling}
Given the knot vector $\mathbf{U}$ and control points derived from the key points (\cref{sec:nurbs_primitives}), we adaptively place isotropic Gaussians along each curve with density proportional to arc length.

We first approximate the arc length $L$ via \cref{eq:arc_length} using $N_L$ uniformly spaced parameter samples, then set the number of Gaussians as $M = \lceil \delta_\text{c} \cdot L \rceil$, where $\delta_\text{c}$ is a user-specified contour density.
The $M$ parameter values $\{u_j\}_{j=1}^{M}$ are distributed uniformly within the valid knot domain $[u_p, u_{m-p}]$.

To evaluate each sample efficiently, we exploit the compact support of B-spline basis functions: at parameter value $u_j$, only the $p{+}1$ basis functions whose support contains $u_j$ are non-zero.
The Gaussian center is therefore computed via a span-local formulation:
\begin{equation}
\boldsymbol{\mu}_j =
\frac{
\sum_{l=0}^{p} N_{i_j-p+l,p}(u_j)\, w_{i_j-p+l}\, \mathbf{P}_{i_j-p+l}
}{
\sum_{l=0}^{p} N_{i_j-p+l,p}(u_j)\, w_{i_j-p+l}
},
\end{equation}
where $i_j$ is the knot span index satisfying $u_{i_j} \le u_j < u_{i_j+1}$.
This reduces evaluation cost from $O(Mn)$ to $O(Mp)$ per curve.
The local stroke width $s_j$ is read from the third coordinate of $\mathbf{C}(u_j)$; higher-order analytical derivatives, used for regularization (\cref{sec:optimization}), are obtained through the same local recursion.

Each sample is assigned an isotropic Gaussian whose scale depends solely on the local stroke width:
\begin{equation}
\sigma_{j,x} = \sigma_{j,y} = s_j / \rho, \qquad \theta_j = 0, \qquad o_j = o,
\end{equation}
where $\rho$ is a global scale ratio controlling overlap between neighboring Gaussians and $o$ is the curve-level opacity from \cref{sec:nurbs_primitives}.
Unlike B\'{e}zier Splatting~\cite{liu2025bzier}, which aligns anisotropic Gaussians to the curve tangent, this isotropic formulation eliminates rotation estimation and avoids visual artifacts---blurry edges at large stroke widths and spikes at high-curvature regions.
Because $M$ scales with arc length, Gaussian centers remain sufficiently dense that circular kernels provide smooth, uniform coverage without directional alignment.

For open curves, we replicate the first and last Gaussian centers $\kappa$ times, where $\kappa$ is a fixed number of times proportional to $\delta_\text{c}$, ensuring opaque, well-defined endpoints without additional parameters.

\subsection{Gaussian Sampling in Regions}
\label{sec:interior_filling}
For closed curves marked as filled, we replace contour Gaussians with a dense set of interior Gaussian primitives whose opacity is modulated by a differentiable signed distance field (SDF).

We first sample $\lceil\delta_\text{b} \cdot L\rceil$ boundary points $\{\mathbf{b}_j\}$ along the closed curve in pixel coordinates, where $\delta_\text{b}$ is a user-specified boundary density and $L$ is the arc length from \cref{sec:gaussian_contour_sampling}.
These boundary points define the polyline against which the SDF is evaluated.

Next, we compute the axis-aligned bounding box of the boundary with symmetric padding proportional to its extent, clipped to the canvas.
A uniform 2D grid at step size~$h$ is placed inside this padded box, yielding $Q = n_x \times n_y$ query points $\{\mathbf{q}_j\}$ in pixel coordinates.
A larger~$h$ produces a coarser grid with fewer Gaussians, trading density for speed; we anneal~$h$ during optimization to accelerate early convergence while preserving final detail (\cref{sec:live}).

For each query point $\mathbf{q}_j$, we generate an isotropic Gaussian primitive:
\begin{equation}
\boldsymbol{\mu}_j = \mathbf{q}_j,
\qquad
\theta_j = 0,
\qquad
\sigma_{j,x} = \sigma_{j,y} = \frac{\eta \cdot h}{\rho},
\end{equation}
where $\eta$ is a coverage factor (we use $\eta{=}1.5$) that ensures sufficient overlap between neighboring Gaussians, and $\rho$ is the global scale ratio from \cref{sec:gaussian_contour_sampling}.
The base opacity is the product of the curve-level opacity $o$ and a sigmoid soft mask derived from the SDF:
\begin{equation}
o_j = o \cdot \mathrm{sigmoid}\!\left(\frac{\gamma}{h} \cdot \mathrm{sdf}(\mathbf{q}_j)\right),
\end{equation}
where $\gamma$ controls the sharpness of the boundary transition, and $\mathrm{sdf}(\mathbf{q}_j)$ returns positive values inside the region and negative values outside, computed with respect to the boundary polyline $\{\mathbf{b}_j\}$.
Detailed algorithmic steps for computing the SDF are provided in the supplementary material.


Each Gaussian inherits its color from the parent curve.
This pixel-space SDF formulation scales to large filled shapes while remaining fully differentiable with respect to boundary point positions.

\subsection{Differentiable Splatting Rendering}
\label{sec:diff_splatting}

Once Gaussian primitives are sampled from NURBS curves, we rasterize them using the tile-based Gaussian splatting rasterizer of GaussianImage~\cite{zhangGaussianImage1000FPS2025}.

The rasterizer sorts the 2D Gaussians by their stacking order (depth) and composites them front-to-back for each pixel.
The rendered color $\mathcal{I}(\mathbf{x}_n)$ at pixel $\mathbf{x}_n$ is obtained by alpha blending the contributions of overlapping Gaussians:
\begin{equation}
\mathcal{I}(\mathbf{x}_n) = \sum_{i \in \mathcal{N}} \mathbf{c}_i \, \alpha_i(\mathbf{x}_n) \prod_{j=1}^{i-1} \bigl(1-\alpha_j(\mathbf{x}_n)\bigr),
\end{equation}
where $\mathcal{N}$ is the depth-ordered set of Gaussians covering pixel $\mathbf{x}_n$, $\mathbf{c}_i$ is the color of the $i$-th Gaussian, and $\alpha_i(\mathbf{x}_n)$ is its opacity at that location.
Any residual transmittance is composited against a predefined background color.

Since all sampled Gaussians are isotropic with $\sigma_{i,x} = \sigma_{i,y} \eqqcolon \sigma_i$ and $\theta_i = 0$ (\cref{sec:gaussian_contour_sampling,sec:interior_filling}), the covariance reduces to $\boldsymbol{\Sigma}_i = \sigma_i^2 \mathbf{I}$ and the per-pixel Gaussian opacity simplifies to
\begin{equation}
\alpha_i(\mathbf{x}_n) = o_i \exp\!\biggl(-\frac{\|\mathbf{x}_n - \boldsymbol{\mu}_i\|^2}{2\,\sigma_i^2}\biggr),
\end{equation}
where $o_i$ is the base opacity and $\boldsymbol{\mu}_i$ is the Gaussian center.
This formulation enables efficient backpropagation of gradients from the image loss through the Gaussian parameters to the NURBS control points and weights.

\subsection{Optimization}
\label{sec:optimization}

The overall objective combines an image-space reconstruction loss with geometric regularization:
\begin{equation}
\mathcal{L}
=
\mathcal{L}_{\text{img}}(\mathcal{I}, \mathcal{I}^{\ast})
+
\lambda_{\text{geo}} \, \mathcal{L}_{\text{geo}},
\end{equation}
where $\mathcal{I}$ and $\mathcal{I}^{\ast}$ are the rendered and target images, $\mathcal{L}_{\text{img}}$ is an image-based loss, and $\mathcal{L}_{\text{geo}}$ enforces regularity on the underlying NURBS curves.
Gradients propagate from pixels through the Gaussian parameters to NURBS control points and weights via the differentiable pipeline above; the specific choice of $\mathcal{L}_{\text{img}}$ is application-dependent.
Below we describe the two regularization terms shared across all applications; additional task-specific losses are introduced in \cref{sec:applications}.

\emph{Derivative Loss},
the squared magnitude of the $r$-th derivative of a parametric curve, is a classical smoothing energy in curve fairing~\cite{pottmannSmoothCurvesTension1990}: the second derivative relates to curvature, and the third captures its rate of change.
We penalize this energy averaged over the parameter domain:
\begin{equation}
\mathcal{L}_{\text{deriv}}^r(\mathbf{C})
=
\frac{1}{u_{\max} - u_{\min}}\int_{u_{\min}}^{u_{\max}}
\left\|
\frac{d^r \mathbf{C}(u)}{d u^r}
\right\|_2^2
\, du,
\end{equation}
where $r$ denotes the derivative order.
We evaluate this integral using analytical derivatives obtained from \cref{eq:curve_derivative} during contour sampling (\cref{sec:gaussian_contour_sampling}), avoiding the numerical errors that arise from finite-difference approximations.
The total loss averages $\mathcal{L}_{\text{deriv}}^r$ over all curves; following Berio \etal~\cite{berioNeuralImageAbstraction2025}, we set $r=3$.

\emph{Bounding Box Loss}
prevents curves from drifting outside the target region. We penalize points that violate the image bounding box:
\begin{equation}
\mathcal{L}_{\text{bbox}}=\sum_i \varphi(\mathbf{b}_{\min} - \mathbf{P}_i) + \varphi(\mathbf{P}_i - \mathbf{b}_{\max}),
\end{equation}
where $\mathbf{b}_{\min}$ and $\mathbf{b}_{\max}$ are the bounding-box corners, $\mathbf{P}_i$ ranges over all sampled curve positions, and $\varphi$ is a soft penalty (ReLU or Softplus) that activates only when a point lies outside the permitted region.

\section{Applications}
\label{sec:applications}
We demonstrate the versatility of NURBS Splatting on three representative vector-graphics tasks: calligraphy reconstruction (\cref{sec:calligraphy}), layer-wise image vectorization (\cref{sec:live}), and neural image abstraction (\cref{sec:neural_abstraction}).

\subsection{Calligraphy Reconstruction}
\label{sec:calligraphy}

In calligraphy, stroke trajectories and their varying widths encode both style and meaning. Recovering these as editable vector strokes from raster images is useful for font design, digital archiving, and robotic handwriting reproduction.

Most calligraphy generation methods produce raster images via GANs~\cite{wuCalliGANStyleStructureaware2020,zengStrokeGANReducingMode2021} or diffusion models~\cite{liaoCalliffusionChineseCalligraphy2023}, discarding stroke structure entirely. Methods that output vector outlines as B\'{e}zier sequences~\cite{wangDeepVecFontSynthesizingHighquality2021,wangDeepVecFontv2ExploitingTransformers2023,thamizharasanVecFusionVectorFont2024} describe glyphs as closed regions, not as strokes with varying width. Classical stroke extraction~\cite{maAutomaticGenerationChinese1995,chenAutomaticStrokeExtraction2017} and the segmentation method of StrokeStyles~\cite{berioStrokeStylesStrokebasedSegmentation2022} do model individual strokes but lack end-to-end differentiable optimization in image space.

Our framework represents each stroke as a NURBS curve with a per-control-point width channel. For initialization, we use the line vectorization method of Magne and Sorkine-Hornung~\cite{magneSingleLineDrawingVectorization2025}: a medial axis is extracted from the binarized image via the Voronoi diagram method~\cite{brandtContinuousSkeletonComputation1992}, a CNN classifies intersections to produce ordered sample points along the axis, and NURBS curves are fit to these points with constant initial width.

The curves are then optimized end-to-end against the raster target with MSE loss and geometric regularization:
\begin{equation}
\label{eq:calligraphy-loss}
  \mathcal{L} = \lambda_{\text{MSE}} \mathcal{L}_{\text{MSE}} + \lambda_{\text{deriv}} \mathcal{L}_{\text{deriv}}^3 + \lambda_{\text{bbox}} \mathcal{L}_{\text{bbox}}.
\end{equation}
Control point positions, NURBS weights, knot intervals, and per-control-point widths are updated jointly, recovering both stroke trajectories and thickness variations. Results are presented in \Cref{sec:calligraphy-exp}.

\subsection{Layer-wise Image Vectorization}
\label{sec:live}

Ma~\etal~\cite{maLayerwiseImageVectorization2022} propose LIVE, a layer-wise vectorization approach that greedily adds closed B\'{e}zier paths one at a time, each initialized at the region of highest reconstruction error and optimized via DiffVG~\cite{liDifferentiableVectorGraphics2020}. A distance-weighted loss modulates per-pixel gradients based on proximity to the newly added shape boundary, concentrating optimization effort where it matters most. We adapt LIVE to our framework, replacing the DiffVG rendering backend with NURBS Splatting while introducing a \emph{grid-step annealing} strategy for filled shapes.

\begin{figure}[t]            
  \centering
  \scriptsize
  \renewcommand{\tabcolsep}{1pt}
  \begin{tabular}{ccccccc}
    \includegraphics[width=0.135\linewidth]{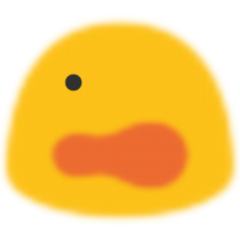} &
    \includegraphics[width=0.135\linewidth]{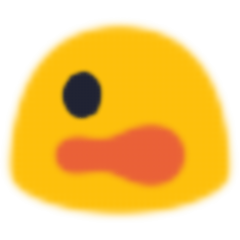} &
    \includegraphics[width=0.135\linewidth]{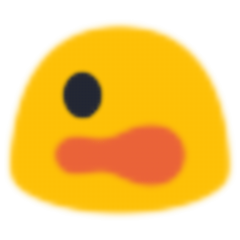} &
    \includegraphics[width=0.135\linewidth]{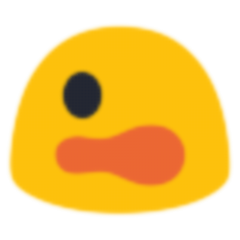} &
    \includegraphics[width=0.135\linewidth]{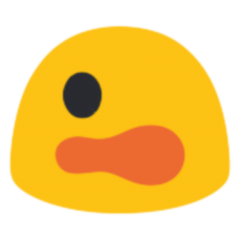} &
    \includegraphics[width=0.135\linewidth]{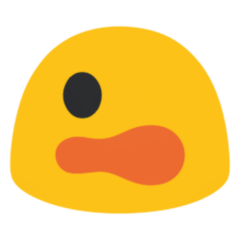} &
    \includegraphics[width=0.135\linewidth]{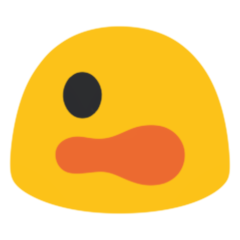} \\
    Iter 0 & Iter 50 & Iter 100 & Iter 150 & Iter 200 & Iter 250 & Iter 299 \\ 
  \end{tabular}
  \caption{Visualization of grid-step annealing. The grid resolution progressively increases, allowing coarse-to-fine optimization.}
  \label{fig:step-anneal}
\end{figure}

To accelerate convergence without sacrificing fidelity, we anneal the fill grid step $h$ from a coarse value ($h_{\max}{=}4$) to a fine one ($h_{\min}{=}1$) using a cosine schedule between iteration fractions $t_b{=}0.1$ and $t_e{=}0.8$:
\begin{equation}
h(t) = h_{\max} + (h_{\min} - h_{\max})\,\frac{1 + \cos(\pi(1 - \phi))}{2}, \quad t/T \in (t_b, t_e),
\end{equation}
where $\phi = (t/T - t_b)/(t_e - t_b)$ and $h$ is clamped outside the range. The coarse grid reduces the number of primitives in the early stages, thereby stabilizing large-scale deformations (\cref{fig:step-anneal}).

The overall loss for each round combines a pixel-wise reconstruction term with geometric regularization:
\begin{equation}
\mathcal{L} = \mathcal{L}_{\text{img}} + \lambda_{\text{deriv}} \mathcal{L}_{\text{deriv}}^3 + \lambda_{\text{xing}} \mathcal{L}_{\text{xing}},
\end{equation}
where $\mathcal{L}_{\text{img}}$ is the distance-weighted MSE loss, $\mathcal{L}_{\text{deriv}}^3$ is the third-order derivative smoothing loss, and $\mathcal{L}_{\text{xing}}$ is a self-crossing penalty adapted from LIVE that penalizes control-point configurations whose consecutive segments reverse winding direction. Results are presented in \Cref{sec:live-exp}.

\subsection{Neural Image Abstraction}
\label{sec:neural_abstraction}

Berio \etal~\cite{berioNeuralImageAbstraction2025} apply long smoothing B-splines to generate abstracted vector images from raster inputs. Because NURBS generalize polynomial B-splines, these applications transfer naturally to our framework while gaining rational-weight flexibility. We demonstrate two representative tasks: single-stroke area filling and diffusion-guided image abstraction.

\begin{figure}[t]
  \centering
  \scriptsize
  \setlength{\tabcolsep}{1pt}
  \begin{tabular}{cccccc}
    \includegraphics[width=0.13\linewidth]{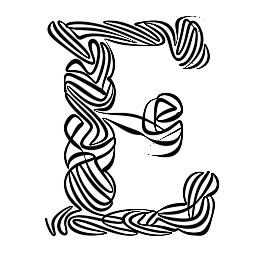} &
    \includegraphics[width=0.13\linewidth]{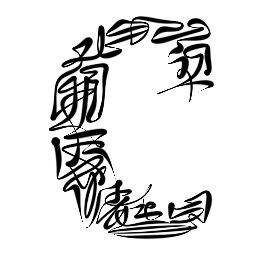} &
    \includegraphics[width=0.13\linewidth]{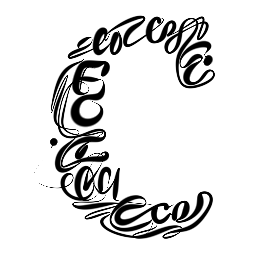} &
    \includegraphics[width=0.13\linewidth]{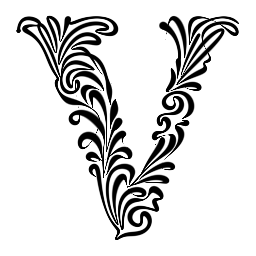} &
    \includegraphics[width=0.13\linewidth]{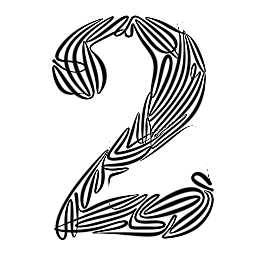} &
    \includegraphics[width=0.13\linewidth]{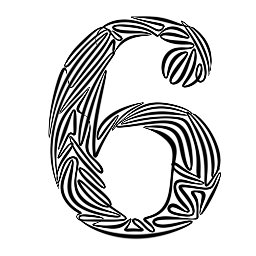} \\
  \end{tabular}
  \caption{Single-stroke area-filling results. Letters are stylized with different style images; digits use coverage loss only (no stylization).}
  \label{fig:area-fill}
\end{figure}

The \emph{Single-Stroke Area Filling} task aims to produce a single NURBS curve that fills a colored shape given its bitmap image.
We initialize a long NURBS curve from a TSP route~\cite{kaplanTSPArt2005} and optimize it with an MSE coverage loss between the rendered and target images. Lowering the target opacity reduces curve density in the interior. To enable semantic stylization, we add a term $\mathcal{L}_{\text{style}}$ based on the patch-wise directional CLIP loss of Kwon and Ye~\cite{kwonCLIPstylerImageStyle2022}, which aligns encoded features of the rendered curves with those of a reference style image. Geometric regularization via $\mathcal{L}_{\text{deriv}}^3$ and $\mathcal{L}_{\text{bbox}}$ enforces smoothness and prevents control-point drift. Representative results appear in \cref{fig:area-fill}.

\begin{figure}[t]
  \centering
  \scriptsize
  \renewcommand{\tabcolsep}{1pt}
  \begin{tabular}{ccccc}
    \begin{tabular}{c} \includegraphics[width=0.19\linewidth]{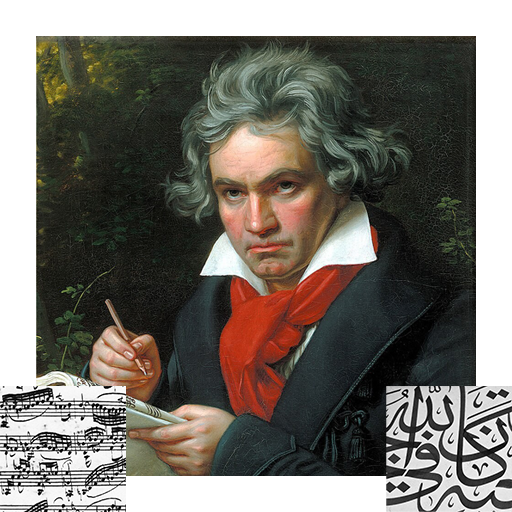} \\  \end{tabular} &
    \begin{tabular}{c} \includegraphics[width=0.182\linewidth]{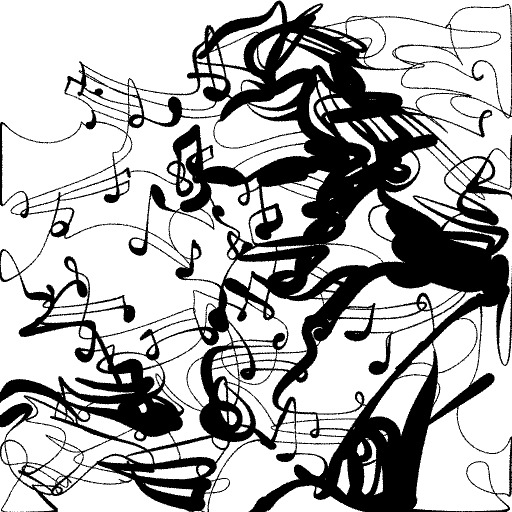} \\ 249.07s \end{tabular} &
    \begin{tabular}{c} \includegraphics[width=0.182\linewidth]{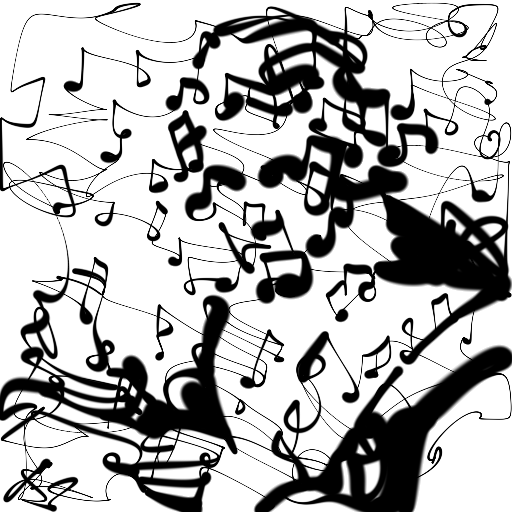} \\ 175.67s \end{tabular} &
    \begin{tabular}{c} \includegraphics[width=0.182\linewidth]{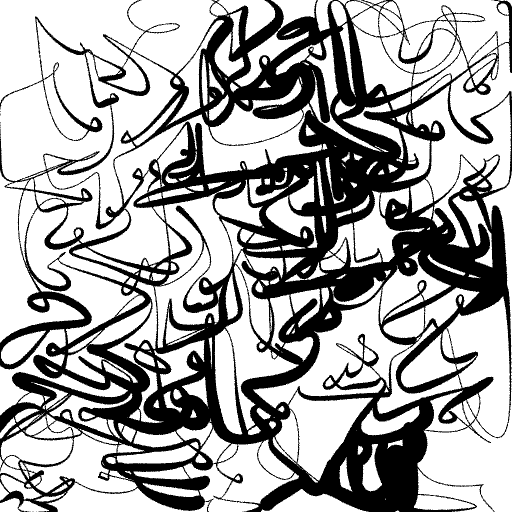} \\ 252.76s \end{tabular} &
    \begin{tabular}{c} \includegraphics[width=0.182\linewidth]{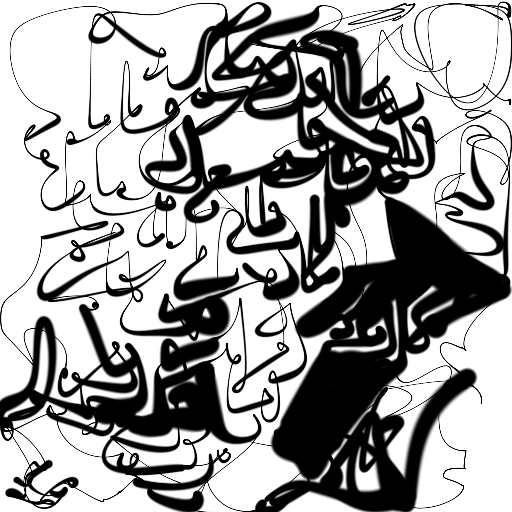} \\ 184.56s \end{tabular} \\
    
    \begin{tabular}{c} \includegraphics[width=0.19\linewidth]{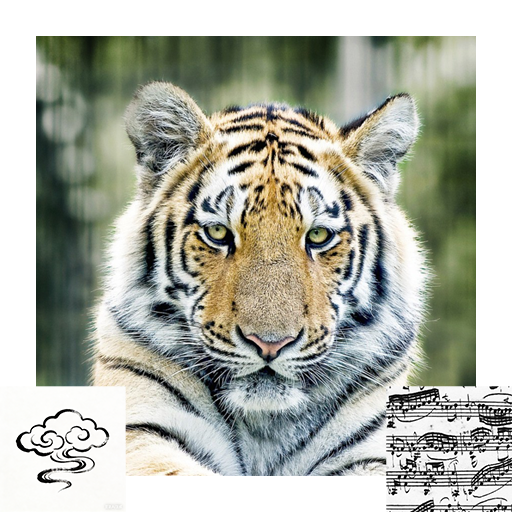} \\  \end{tabular} &
    \begin{tabular}{c} \includegraphics[width=0.182\linewidth]{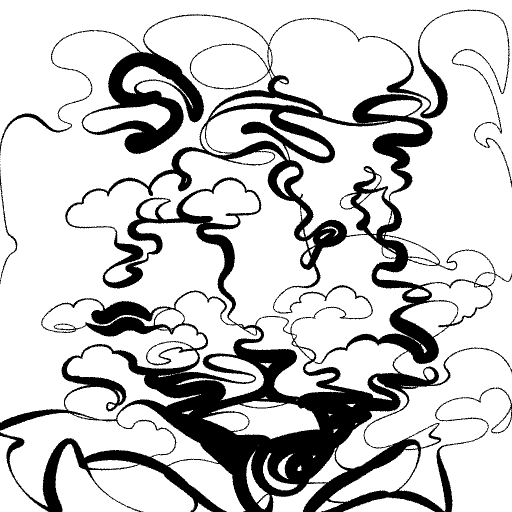} \\ 213.36s \end{tabular} &
    \begin{tabular}{c} \includegraphics[width=0.182\linewidth]{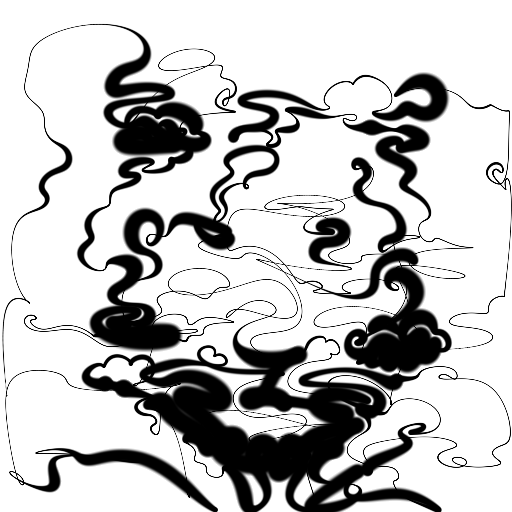} \\ 201.84s \end{tabular} &
    \begin{tabular}{c} \includegraphics[width=0.182\linewidth]{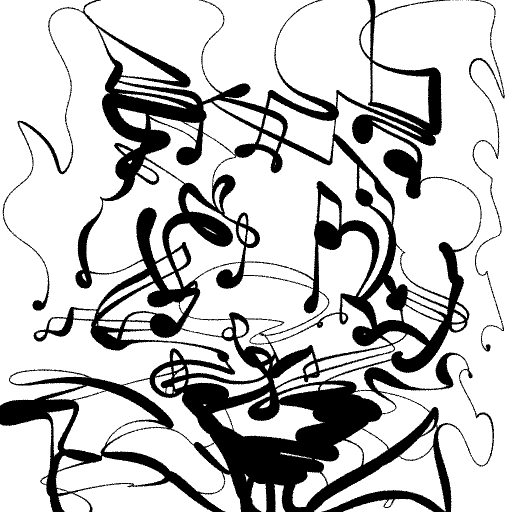} \\ 217.14s \end{tabular} &
    \begin{tabular}{c} \includegraphics[width=0.182\linewidth]{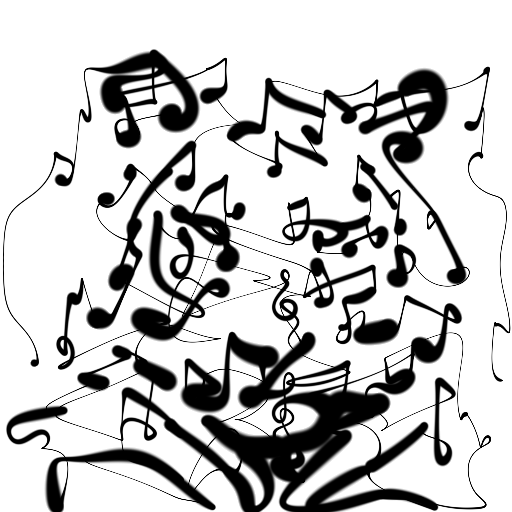} \\ 176.93s \end{tabular} \\
    
    Input & Berio \etal~\cite{berioNeuralImageAbstraction2025} & Ours & Berio \etal~\cite{berioNeuralImageAbstraction2025} & Ours \\
  \end{tabular}
  \caption{Image abstraction with semantic stylization. Style images are inset in the bottom corners of the inputs. Our method matches stylization quality while running $1.27\times$ faster (single NVIDIA RTX 4090, 512$\times$512, 300 iterations).}
  \label{fig:abstraction}
\end{figure}

For the \emph{Single-Stroke Image Abstraction} task, we couple our renderer with a diffusion-based Score Distillation Sampling (SDS) pipeline.
Following Berio \etal~\cite{berioNeuralImageAbstraction2025}, the pipeline uses ControlNet~\cite{zhangAddingConditionalControl2023} with Canny edge detection and IP-Adapter~\cite{yeIPAdapterTextCompatible2023}, guided by the text prompt ``A black and white ink drawing.'' The SDS gradient with respect to NURBS parameters $\boldsymbol{\psi}$ is
\begin{equation}
\nabla_{\boldsymbol{\psi}} \mathcal{L}_{\text{SDS}}(\boldsymbol{\psi})
=
\mathbb{E}_{t}
\left[
w(t)
\left(\hat{\boldsymbol\epsilon}(x_t, t, y) - \boldsymbol\epsilon\right)
\frac{\partial g(\boldsymbol{\psi})}{\partial \boldsymbol{\psi}}\right],
\end{equation}
where $\hat{\boldsymbol\epsilon}(x_t, t, y)$ is the predicted denoising direction of the frozen diffusion model for a latent $x_t$ at time step $t$,
$\boldsymbol\epsilon$ is the sampled noise,
and $w(t)$ is a weighting function dependent on $t$.
The final loss combines SDS with geometric regularization $\mathcal{L}_{\text{deriv}}^3$ and $\mathcal{L}_{\text{bbox}}$. As shown in \cref{fig:abstraction}, our method achieves $1.27\times$ faster optimization than Berio \etal~\cite{berioNeuralImageAbstraction2025} at comparable stylization quality---even though differentiable rendering is not the primary bottleneck of the SDS pipeline and NURBS evaluation is more complex than polynomial B-spline evaluation.

\section{Experiments}
We implement our NURBS Splatting framework in PyTorch.
We show general comparisons on rendering efficiency, quality, and editability in \cref{sec:chore};
experiments for our calligraphy reconstruction application are detailed in \cref{sec:calligraphy-exp};
experiments for LIVE are in \cref{sec:live-exp}.
All experiments are run on a single NVIDIA RTX 4090.
Code and dataset are available at \url{https://github.com/AnicoderAndy/nurbs-splatting}.

\subsection{Efficiency, Quality, and Editability}
\label{sec:chore}
Our method shows improved efficiency compared to Berio \etal~\cite{berioNeuralImageAbstraction2025} when the number of control points scales, as it avoids matrix multiplication to convert the splines. However, the superiority is not seen when rendering with simple geometric primitives, comparing our method with Berio \etal~\cite{berioNeuralImageAbstraction2025} and B\'{e}zier Splatting~(BS)~\cite{liu2025bzier}. Details are provided in the supplementary material.

\begin{figure}[t]
  \centering
  \scriptsize
  \setlength{\tabcolsep}{1pt}
  \begin{tabular}{cccccc}
    \includegraphics[width=0.152\linewidth]{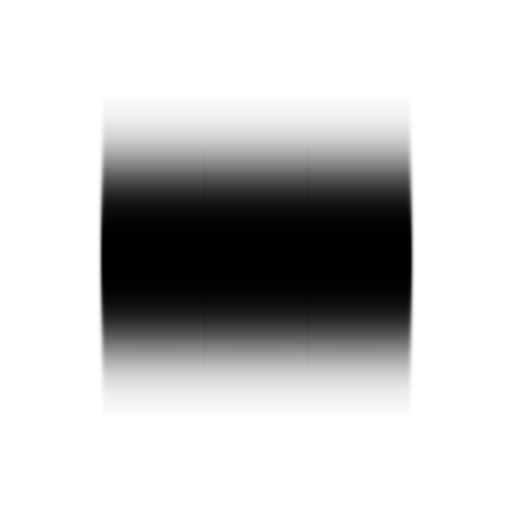} &
    \includegraphics[width=0.152\linewidth]{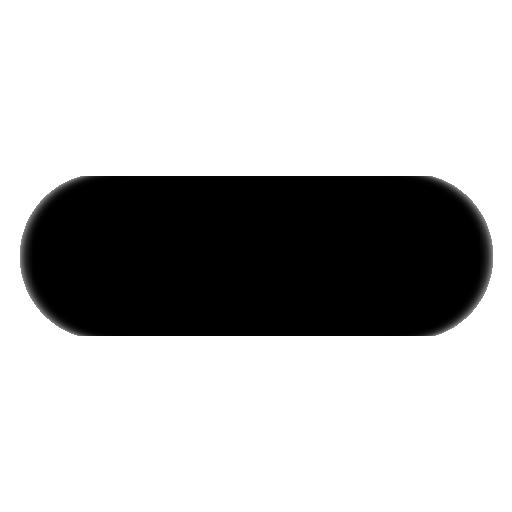} &
    \includegraphics[width=0.152\linewidth]{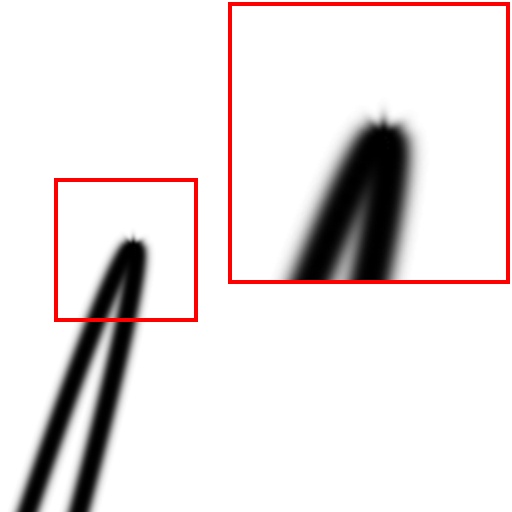} &
    \includegraphics[width=0.152\linewidth]{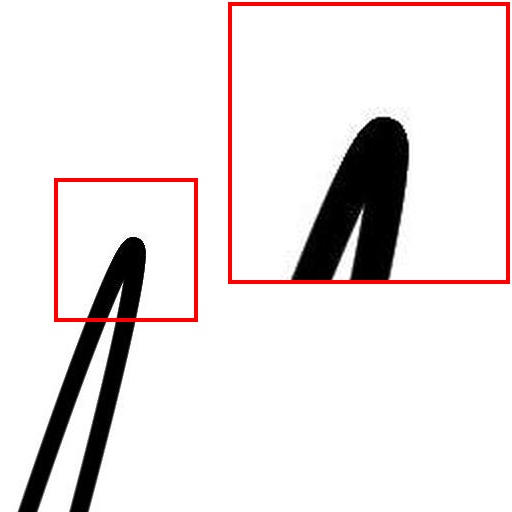} &
    \includegraphics[width=0.152\linewidth]{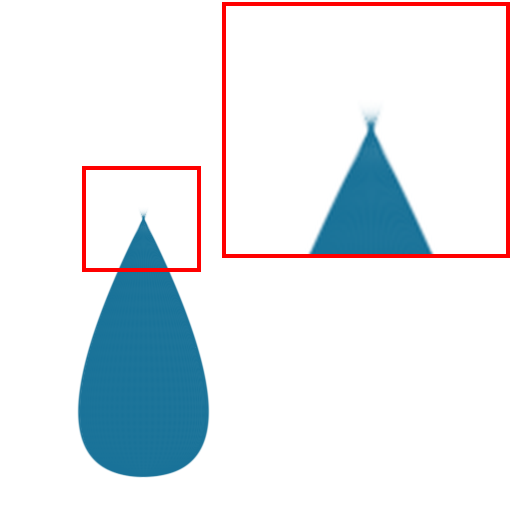} &
    \includegraphics[width=0.152\linewidth]{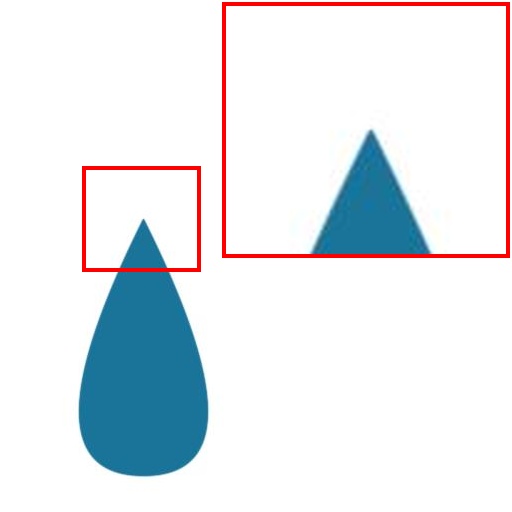} \\
    BS & Ours & BS & Ours & BS & Ours
  \end{tabular}
  \caption{Rendering comparison on three diagnostic cases. B\'{e}zier Splatting (BS) shows blurred edges on thick strokes (left pair), spike artifacts near high-curvature regions (middle pair, local crop), and spikes at filled-region junctions (right pair). B\'{e}zier splines are equivalently converted to NURBS for fair geometric comparison.}
  \label{fig:render-quality}
\end{figure}

As for rendering quality, we qualitatively compare our method with BS which uses anisotropic Gaussian sampling on three diagnostic cases (\cref{fig:render-quality}).
For thick strokes, BS exhibits severely blurred edges.
In the high-curvature case, BS produces spike artifacts due to the discontinuity and instability of Gaussian rotation evaluation via \texttt{atan2}.
In the filled-region case, BS generates spikes at region junctions caused by incorrect curve interpolation at the shared endpoints of paired B\'{e}zier curves.
Our method resolves all three artifacts through purely isotropic Gaussian sampling, consistently yielding sharp boundaries and smooth geometry.

A major advantage of vector graphics is their editability.
The control points of optimized NURBS curves from our method provide a clean structure allowing manual adjustment. We present a visualization in the supplementary material.

\subsection{Calligraphy Reconstruction}
\label{sec:calligraphy-exp}
We construct a calligraphy benchmark of 192 characters: all 92 Japanese Kana and the first 100 characters from General Standard Chinese Characters~\cite{ministryofeducationofthepeoplesrepublicofchinaTableGeneralStandard2013}. Each character is rasterized from existing fonts~\cite{keikan&midnightgardenYoppaFudeFont,foundertypefoundryFounderOuyangXun} to $512{\times}512$ images.

Since, to the best of our knowledge, no existing method performs stroke-level reconstruction of calligraphy, we implement a baseline based on Berio \etal~\cite{berioNeuralImageAbstraction2025}, which optimizes uniform B-splines.
Both methods use identical initialization via medial-axis extraction~\cite{magneSingleLineDrawingVectorization2025}.
We evaluate reconstruction quality using standard pixel-level measures (MSE, PSNR, SSIM) and geometry-aware metrics (symmetric Hausdorff distance and F1 score on binarized stroke masks).

\begin{table}[t]
  \centering
  \setlength{\tabcolsep}{3pt}
  \caption{Quantitative comparison of calligraphy reconstruction quality and runtime. The baseline is Berio \etal~\cite{berioNeuralImageAbstraction2025}. Best and second-best results are highlighted in \textbf{bold} and \underline{underline}, respectively.}
  \begin{tabular}{l c c c c c c}
    \toprule
    Method & MSE$\downarrow$ & PSNR$\uparrow$ & SSIM$\uparrow$ & Hausdorff$\downarrow$ & F1$\uparrow$ & Opt. Time (s) \\
    \midrule
    Baseline & 0.0057 & 24.72 & \underline{0.9813} & 15.26 & 0.9642 & 10.65 \\
    Ours & \textbf{0.0038} & \textbf{26.32} & 0.9793 & \textbf{10.69} & \underline{0.9741} & 8.73 \\
    Ours ($\delta_\text{c}{=}30$) & \underline{0.0039} & \underline{26.24} & \textbf{0.9824} & 11.07 & \textbf{0.9743} & 9.34 \\
    \midrule
    Ours w/o weights & 0.0046 & 25.63 & 0.9777 & 11.70 & 0.9701 & 8.98 \\
    Ours w/o knots & 0.0040 & 26.07 & 0.9789 & \underline{10.98} & 0.9731 & \underline{5.63} \\
    Ours w/o w \& k & 0.0042 & 25.92 & 0.9785 & 11.34 & 0.9721 & \textbf{5.12} \\
    \bottomrule
  \end{tabular}
  \label{tab:calligraphy-metrics}
\end{table}
\begin{figure}[t]
  \centering
  \setlength{\tabcolsep}{1pt}
  \begin{tabular}{cccccc}
    \includegraphics[width=0.16\linewidth]{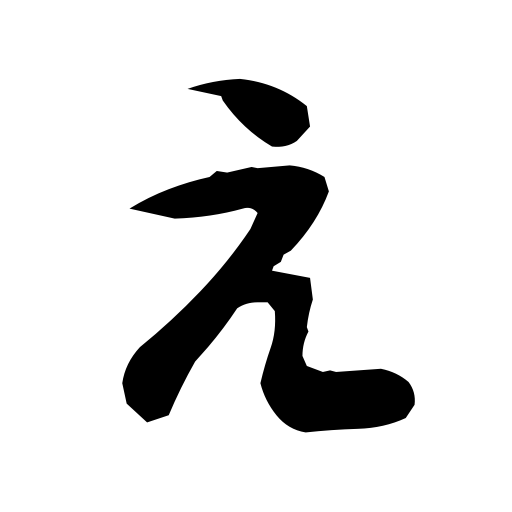} &
    \includegraphics[width=0.16\linewidth]{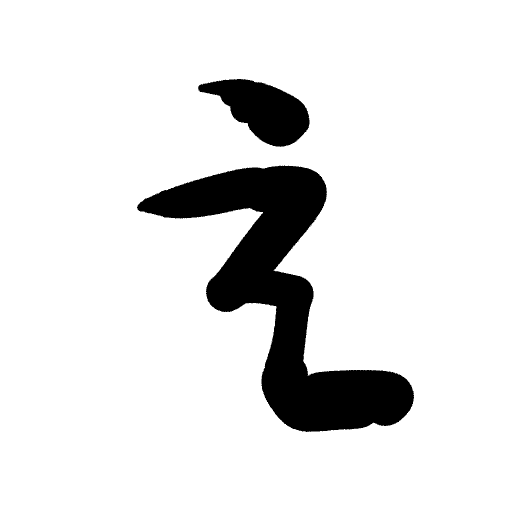} &
    \includegraphics[width=0.16\linewidth]{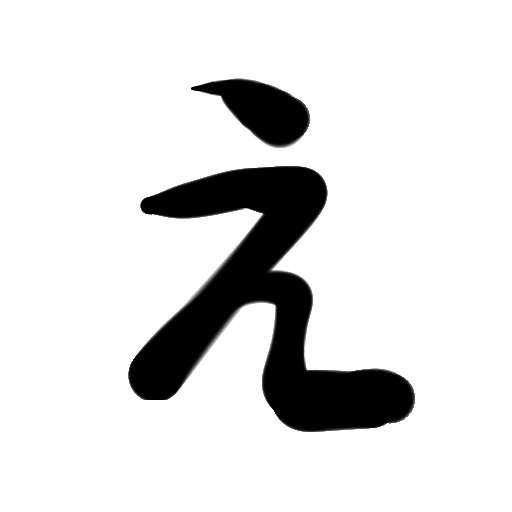} &
    \includegraphics[width=0.16\linewidth]{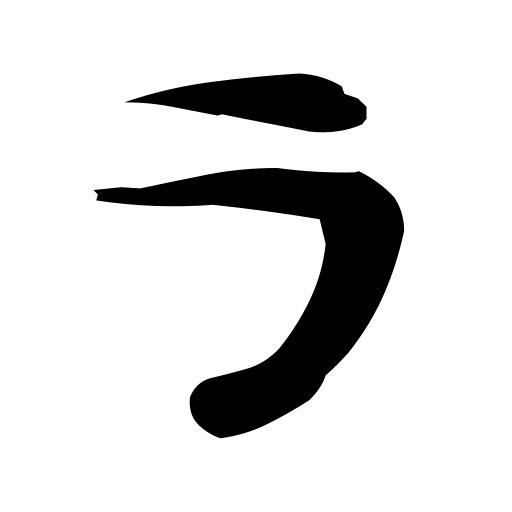} &
    \includegraphics[width=0.16\linewidth]{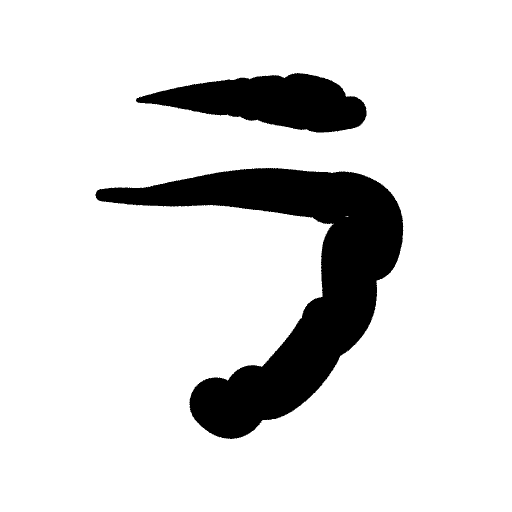} &
    \includegraphics[width=0.16\linewidth]{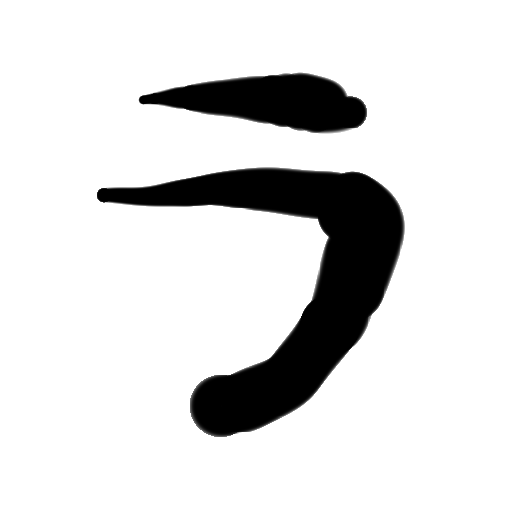} \\
    \includegraphics[width=0.16\linewidth]{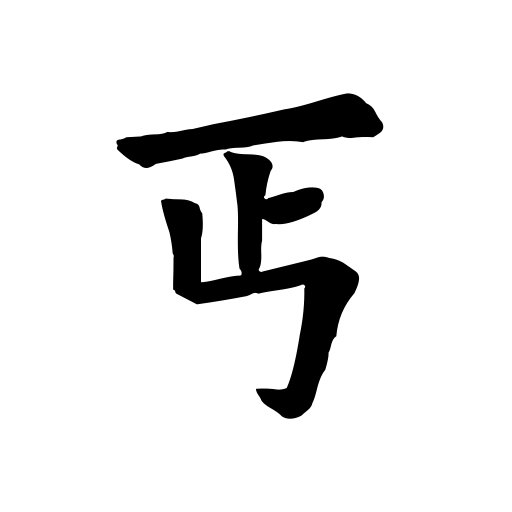} &
    \includegraphics[width=0.16\linewidth]{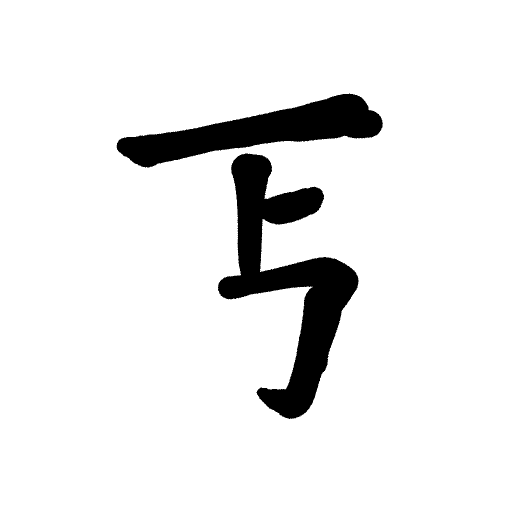} &
    \includegraphics[width=0.16\linewidth]{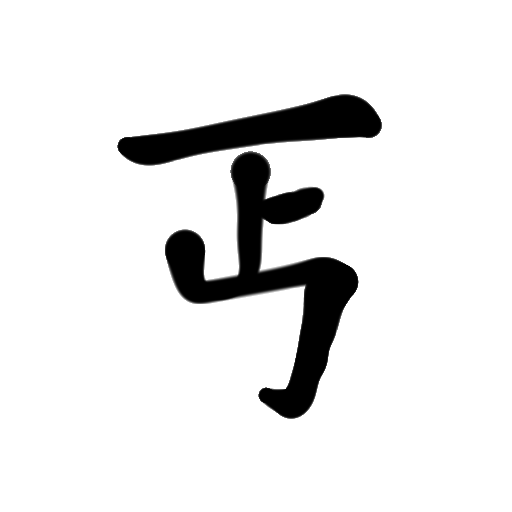} &
    \includegraphics[width=0.16\linewidth]{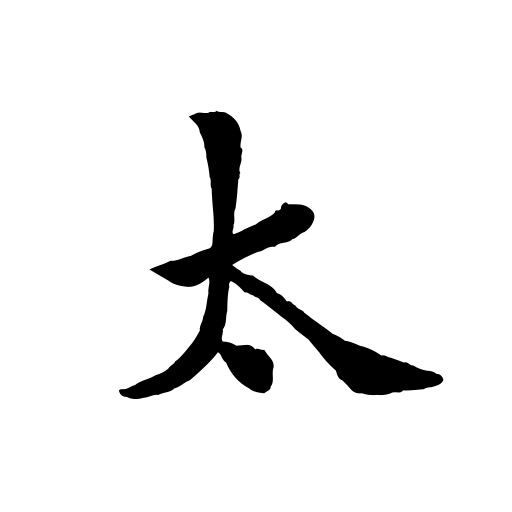} &
    \includegraphics[width=0.16\linewidth]{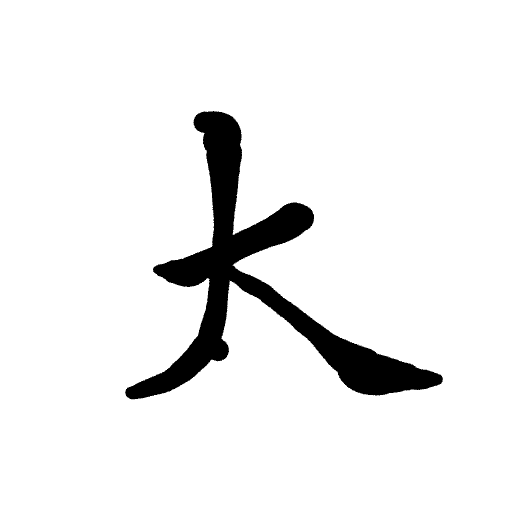} &
    \includegraphics[width=0.16\linewidth]{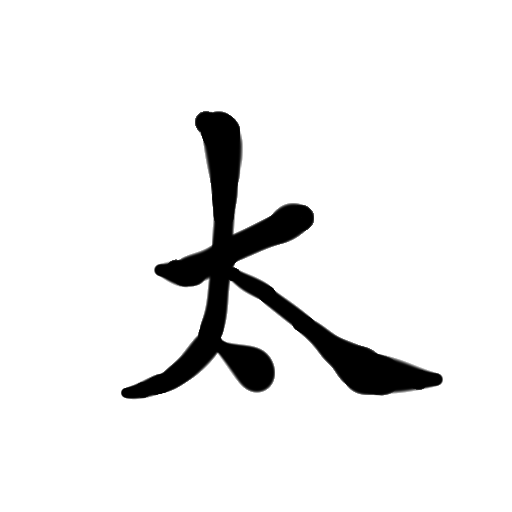} \\
    GT & Baseline & Ours & GT & Baseline & Ours \\
  \end{tabular}
  \caption{Qualitative comparison of calligraphy reconstruction. Top row: Japanese characters. Bottom row: Chinese characters. Our method better preserves stroke trajectories and width variations compared to the baseline~\cite{berioNeuralImageAbstraction2025}.}
  \label{fig:calligraphy-comparison}
\end{figure}

\Cref{tab:calligraphy-metrics} shows the quantitative results averaged over 191 characters.
At the contour density $\delta_\text{c}{=}18$, our full method reduces MSE by $33\%$, improves PSNR by $+1.6$\,dB, and lowers the Hausdorff distance by $30\%$, while running $1.2{\times}$ faster.
The speed gain stems from the avoidance of the matrix multiplication required to convert B-splines into Bézier curves.
Gains are most pronounced on Japanese Kana ($+$2.10\,dB PSNR), where frequent sharp bends and hooks benefit from non-uniform knot spacing that concentrates basis-function support in high-curvature regions.
\Cref{fig:calligraphy-comparison} presents qualitative comparisons on representative characters, showing that our method more faithfully preserves stroke trajectories and width variation than the baseline.

The default configuration drops slightly in SSIM because isotropic Gaussians produce softer boundary transitions than the analytic edge rendering of DiffVG.
Increasing the sampling density to $\delta_\text{c}{=}30$ sharpens these boundaries, recovering the best SSIM ($\textbf{0.9824}$) and achieving the highest F1 ($\textbf{0.9743}$) at only a modest increase in optimization time.

Comparing only uniform polynomial B-splines---the Baseline against our ``w/o w~\&~k'' variant---isolates the effect of the rendering backend: splatting yields a $2.1{\times}$ speedup ($5.12$\,s vs.\ $10.65$\,s) while simultaneously improving PSNR by $+1.2$\,dB and Hausdorff by $26\%$. This confirms that the Gaussian rasterization pipeline is consistently beneficial even without NURBS-specific parameters.

\emph{Ablation studies} (bottom three rows of \cref{tab:calligraphy-metrics}) isolate the individual contributions of learnable weights and knots.
Removing rational weights (``w/o weights'') incurs the largest single-component degradation ($-0.69$\,dB PSNR, $+1.01$\,px Hausdorff relative to the full model), indicating that per-control-point weighting is especially valuable for modulating local curvature.
Removing knots (``w/o knots'') has a smaller effect on reconstruction quality but provides the largest time saving ($5.63$\,s).
Jointly dropping both (``w/o w~\&~k'') still comfortably outperforms the baseline on all metrics except SSIM, while reducing optimization time to roughly half that of the baseline ($5.12$\,s)---offering a strong quality--speed trade-off for applications where rational parameterization is unnecessary.
Importantly, the benefits of weights and knots are complementary: the full model improves $+0.25$\,dB over the next-best single-ablation variant (``w/o knots''), confirming that rational weights and non-uniform knots address distinct geometric limitations.

\subsection{Layer-wise Image Vectorization}
\label{sec:live-exp}

Since the original dataset used in LIVE~\cite{maLayerwiseImageVectorization2022} is not publicly available, we construct a new evaluation benchmark using Noto~Emoji~v1~\cite{NotoEmoji2026}, following their protocol. We select 100 representative emojis and rasterize them at a resolution of 480$\times$480. We compare our method against the original LIVE~\cite{maLayerwiseImageVectorization2022}, which uses DiffVG~\cite{liDifferentiableVectorGraphics2020} as its rendering backend.

In each round, both methods initialize a new closed path with 12 key points as a circle and optimize for 300 iterations; 10 rounds are run in total, yielding 10 paths per image.
We report standard reconstruction metrics: MSE, SSIM, and LPIPS~\cite{zhangUnreasonableEffectivenessDeep2018}, and measure the total optimization time per image.

\begin{table}[t]
  \centering
  \setlength{\tabcolsep}{5pt}
  \caption{Quantitative comparison on layer-wise image vectorization (100 Noto Emoji images at $480{\times}480$). Best and second-best results are in \textbf{bold} and \underline{underline}.}
  \begin{tabular*}{0.8\linewidth}{l @{\extracolsep{\fill}} c c c c}
    \toprule
    Method & MSE$\downarrow$ & SSIM$\uparrow$ & LPIPS$\downarrow$ & Time (s)$\downarrow$ \\
    \midrule
    Baseline & 0.0025 & 0.9633 & 0.0576 & 590 \\
    Ours & \underline{0.0017} & \underline{0.9803} & \underline{0.0339} & \underline{410} \\
    Ours w/o step & \textbf{0.0015} & \textbf{0.9839} & \textbf{0.0334} & 527 \\
    Ours w/o w \& k & 0.0018 & 0.9795 & 0.0348 & \textbf{354} \\
    \bottomrule
  \end{tabular*}
  \label{tab:live-quantitative}
\end{table}

\Cref{tab:live-quantitative} summarizes quantitative results averaged over 100 emojis.
All our variants substantially outperform the LIVE baseline on every metric while also requiring less optimization time.

Our full method with the grid-step annealing strategy reduces MSE by $32\%$, improves SSIM by $+0.017$, and lowers LPIPS by $41\%$, while achieving a $1.4{\times}$ speedup.
Disabling step annealing (``w/o step'') further improves reconstruction quality at the cost of longer optimization ($527$\,s), yet remains $11\%$ faster than LIVE, indicating that our optimization framework remains computationally efficient even without the annealing strategy.
Removing rational weights and knots (``w/o w~\&~k'') still outperforms the baseline on all metrics---reducing MSE by $28\%$, improving SSIM by $+0.016$, and lowering LPIPS by $40\%$---while achieving the fastest optimization at $354$\,s ($1.7{\times}$ speedup), albeit with a smaller quality margin than the full model.

\begin{figure}[t]
  \centering
  \setlength{\tabcolsep}{1pt}
  \begin{tabular}{cccccc}
    \includegraphics[width=0.15\linewidth]{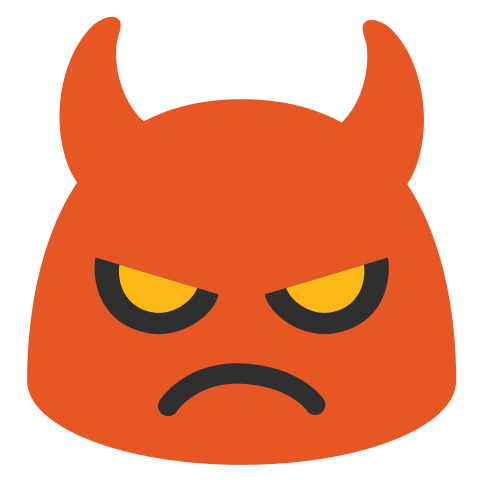} &
    \includegraphics[width=0.15\linewidth]{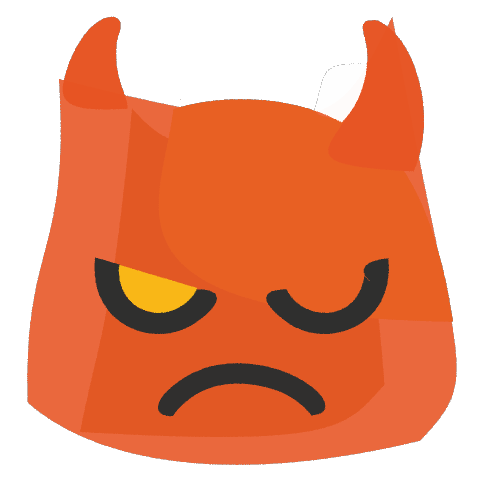} &
    \includegraphics[width=0.15\linewidth]{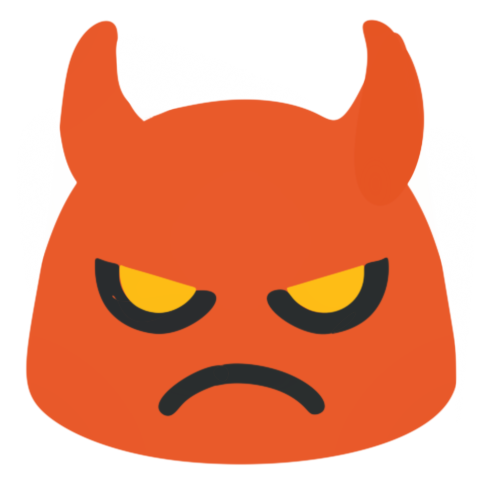} &
    \includegraphics[width=0.15\linewidth]{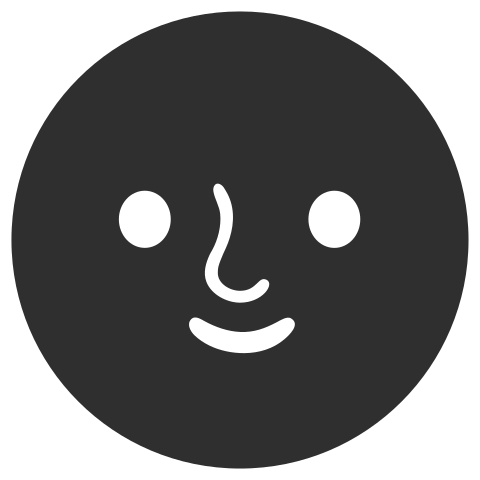} &
    \includegraphics[width=0.15\linewidth]{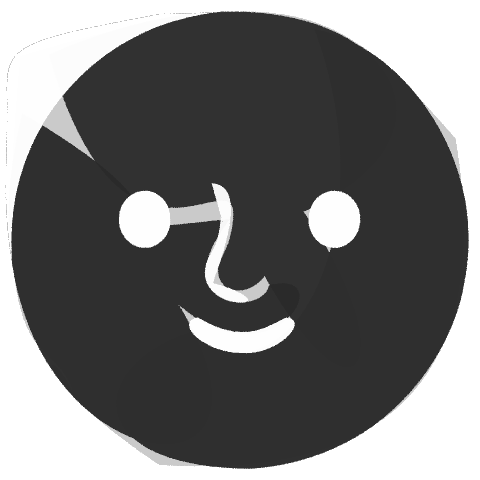} &
    \includegraphics[width=0.15\linewidth]{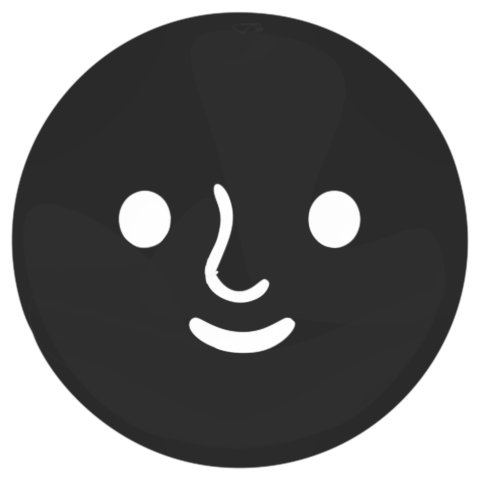} \\
    \includegraphics[width=0.15\linewidth]{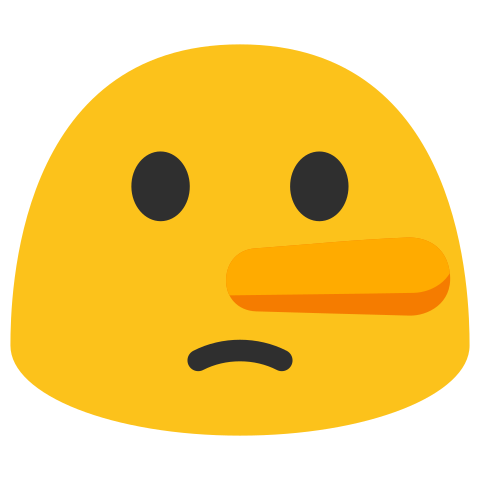} &
    \includegraphics[width=0.15\linewidth]{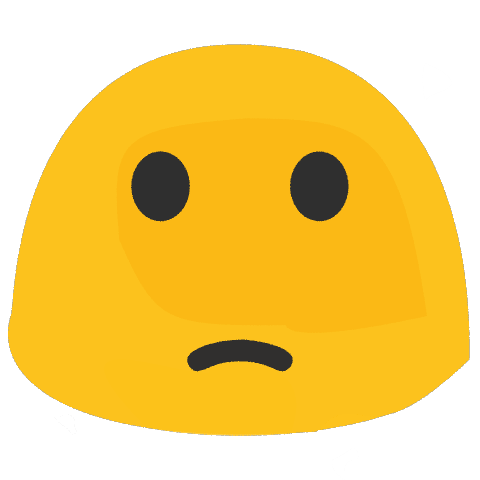} &
    \includegraphics[width=0.15\linewidth]{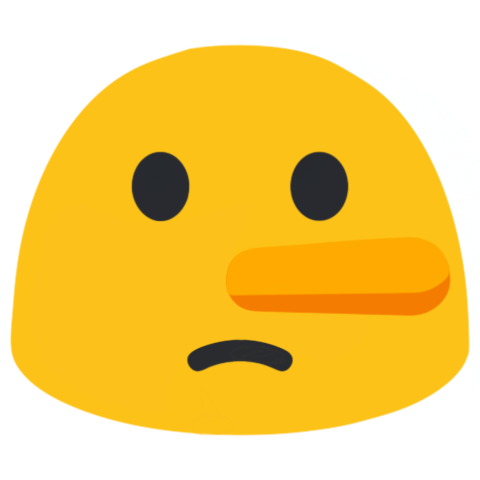} &
    \includegraphics[width=0.15\linewidth]{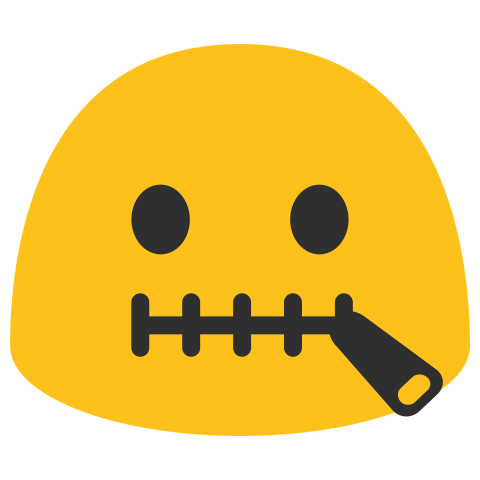} &
    \includegraphics[width=0.15\linewidth]{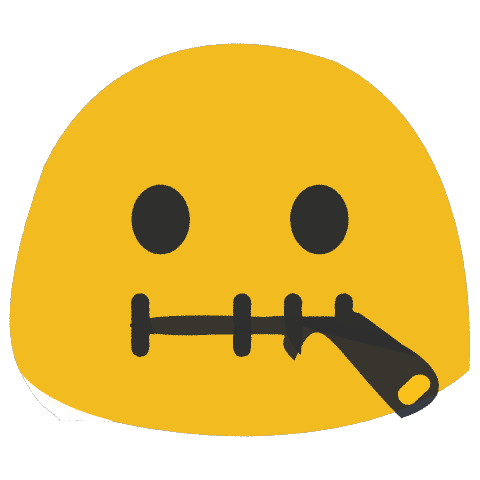} &
    \includegraphics[width=0.15\linewidth]{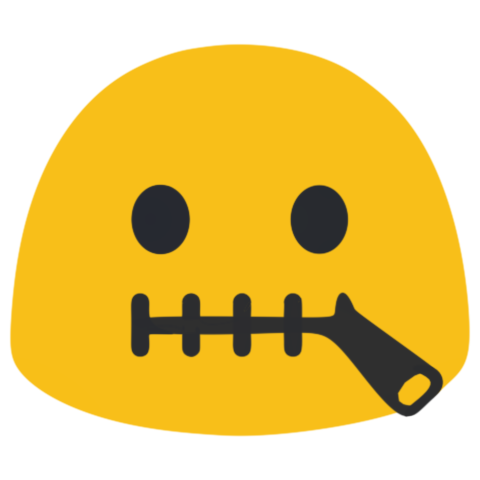} \\
    GT & Baseline & Ours & GT & Baseline & Ours \\
  \end{tabular}
  \caption{Qualitative comparison of vector reconstruction. Our method faithfully reconstructs complex facial expressions and fine details where the DiffVG-based baseline tends to fail.}
  \label{fig:live-qualitative}
\end{figure}

Qualitatively, \cref{fig:live-qualitative} demonstrates that our method is more robust to complex topologies, particularly for thin interleaving structures in facial expressions, where the DiffVG-based LIVE baseline often exhibits artifacts such as missing details and geometric distortions.

\section{Conclusion}
We presented NURBS Splatting, the first differentiable rendering framework capable of optimizing rational B-splines in 2D image space.
By representing curves and filled regions as continuous Gaussian fields, our method unifies vector graphics rendering with gradient-based optimization, allowing joint refinement of control points, rational weights, and non-uniform knot vectors.
Experiments on calligraphy reconstruction and image vectorization demonstrate consistent improvements over polynomial baselines, achieving higher reconstruction fidelity.
Crucially, the ability to optimize rational weights introduces geometric flexibility beyond prior differentiable vector graphics methods, narrowing the gap between learning-based reconstruction and CAD-standard representations.

Our current region-filling strategy relies on a heuristic SDF-based grid, which provides limited advantages over scanline methods for regular geometries.
In addition, although splatting enables efficient rendering, the number of Gaussian primitives can grow with curve complexity, potentially increasing memory usage for highly detailed illustrations compared to implicit representations.
Future work will focus on extending the framework to 3D NURBS surfaces to support native CAD geometry, developing a more efficient analytic region-filling algorithm, and exploring integration with generative models to synthesize editable vector designs directly from text or images.

\section*{Acknowledgements}
We thank the anonymous reviewers for their valuable feedback and suggestions.
This work was funded by the National Natural Science Foundation of China, No.~62076090; the Huxiang Youth Talent Support Program, Hunan Province, China, No.~2020RC3014; and the Natural Science Foundation of Hunan Province, China, No.~2022JJ30173.

\bibliographystyle{splncs04}
\bibliography{references}

@String(CVPR  = {IEEE Conf. Comput. Vis. Pattern Recog.})

@String(ICCV  = {Int. Conf. Comput. Vis.})

@String(ECCV  = {Eur. Conf. Comput. Vis.})

@String(AAAI  = {AAAI})

@book{10.5555/265261,
  title     = {The {{NURBS}} Book (2nd Ed.)},
  author    = {Piegl, Les and Tiller, Wayne},
  year      = 1997,
  publisher = {Springer-Verlag},
  address   = {Berlin, Heidelberg},
  isbn      = {3-540-61545-8}
}

@article{berioNeuralImageAbstraction2025,
  title   = {Neural {{Image Abstraction}} Using {{Long Smoothing B-Splines}}},
  author  = {Berio, Daniel and Stroh, Michael and Calinon, Sylvain and Fol Leymarie, Frederic and Deussen, Oliver and Shamir, Ariel},
  year    = 2025,
  month   = dec,
  journal = {ACM Trans. Graph.},
  volume  = {44},
  number  = {6},
  pages   = {225:1--225:11},
  issn    = {0730-0301, 1557-7368},
  doi     = {10.1145/3763345},
  urldate = {2025-12-29},
  langid  = {english}
}

@article{berioStrokeStylesStrokebasedSegmentation2022,
  title      = {{{StrokeStyles}}: {{Stroke-based Segmentation}} and {{Stylization}} of {{Fonts}}},
  shorttitle = {{{StrokeStyles}}},
  author     = {Berio, Daniel and Leymarie, Frederic Fol and Asente, Paul and Echevarria, Jose},
  year       = 2022,
  month      = apr,
  journal    = {ACM Trans. Graph.},
  volume     = {41},
  number     = {3},
  pages      = {28:1--28:21},
  issn       = {0730-0301},
  doi        = {10.1145/3505246},
  urldate    = {2026-02-17},
  langid     = {american}
}

@article{brandtContinuousSkeletonComputation1992,
  title   = {Continuous Skeleton Computation by {{Voronoi}} Diagram},
  author  = {Brandt, Jonathan W. and Algazi, V. Ralph},
  year    = 1992,
  month   = may,
  journal = {CVGIP: Image Understanding},
  volume  = {55},
  number  = {3},
  pages   = {329--338},
  issn    = {1049-9660},
  doi     = {10.1016/1049-9660(92)90030-7},
  urldate = {2026-02-02}
}

@article{chakrabortyImageVectorizationGradient2025,
  title     = {Image {{Vectorization}} via {{Gradient Reconstruction}}},
  author    = {Chakraborty, Souymodip and Batra, Vineet and Phogat, Ankit and Jain, Vishwas and Ranawat, Jaswant Singh and Dhingra, Sumit and Wampler, Kevin and Fisher, Matthew and Luk{\'a}{\v c}, Michal},
  year      = 2025,
  month     = may,
  journal   = {Comput. Graph. Forum},
  volume    = {44},
  number    = {2},
  publisher = {Wiley},
  issn      = {0167-7055, 1467-8659},
  doi       = {10.1111/cgf.70055},
  urldate   = {2025-07-23},
  copyright = {http://onlinelibrary.wiley.com/termsAndConditions\#vor},
  langid    = {english}
}

@inproceedings{chenAutomaticStrokeExtraction2017,
  title     = {An Automatic Stroke Extraction Method Using Manifold Learning},
  booktitle = {Proceedings of the {{European Association}} for {{Computer Graphics}}: {{Short Papers}}},
  author    = {Chen, Xudong and Lian, Zhouhui and Tang, Yingmin and Xiao, Jianguo},
  year      = 2017,
  month     = apr,
  series    = {{{EG}} '17},
  pages     = {65--68},
  editor    = {Adrien Peytavie and Carles Bosch},
  publisher = {Eurographics Association},
  address   = {Goslar, DEU},
  doi       = {10.2312/egsh.20171016},
  urldate   = {2026-02-17}
}

@article{coxNumericalEvaluationBSplines1972,
  title   = {The {{Numerical Evaluation}} of {{B-Splines}}},
  author  = {Cox, M. G.},
  year    = 1972,
  month   = oct,
  journal = {IMA Journal of Applied Mathematics},
  volume  = {10},
  number  = {2},
  pages   = {134--149},
  issn    = {0272-4960},
  doi     = {10.1093/imamat/10.2.134},
  urldate = {2026-02-26}
}

@article{deboorCalculatingBsplines1972,
  title   = {On Calculating with {{B-splines}}},
  author  = {{de Boor}, Carl},
  year    = 1972,
  month   = jul,
  journal = {Journal of Approximation Theory},
  volume  = {6},
  number  = {1},
  pages   = {50--62},
  issn    = {0021-9045},
  doi     = {10.1016/0021-9045(72)90080-9},
  urldate = {2026-02-26}
}

@article{devaprasadNURBSDiffDifferentiableProgramming2022,
  title      = {{{NURBS-Diff}}: {{A Differentiable Programming Module}} for {{NURBS}}},
  shorttitle = {{{NURBS-Diff}}},
  author     = {Deva Prasad, Anjana and Balu, Aditya and Shah, Harshil and Sarkar, Soumik and Hegde, Chinmay and Krishnamurthy, Adarsh},
  year       = 2022,
  month      = may,
  journal    = {Computer-Aided Design},
  volume     = {146},
  pages      = {103199},
  issn       = {00104485},
  doi        = {10.1016/j.cad.2022.103199},
  urldate    = {2026-01-30},
  langid     = {english}
}

@misc{fanNeuroNURBSLearningEfficient2024,
  title         = {{{NeuroNURBS}}: {{Learning Efficient Surface Representations}} for {{3D Solids}}},
  shorttitle    = {{{NeuroNURBS}}},
  author        = {Fan, Jiajie and Gholami, Babak and B{\"a}ck, Thomas and Wang, Hao},
  year          = 2024,
  month         = nov,
  number        = {arXiv:2411.10848},
  eprint        = {2411.10848},
  primaryclass  = {cs},
  publisher     = {arXiv},
  doi           = {10.48550/arXiv.2411.10848},
  urldate       = {2026-02-26},
  archiveprefix = {arXiv}
}

@book{farinCurvesSurfacesCAGD2007,
  title      = {Curves and {{Surfaces}} for {{CAGD}}: {{A Practical Guide}}},
  shorttitle = {Curves and {{Surfaces}} for {{CAGD}}},
  author     = {Farin, Gerald},
  year       = 2007,
  month      = sep,
  publisher  = {Morgan Kaufmann},
  doi        = {10.1016/B978-1-55860-737-8.X5000-5},
  urldate    = {2026-02-27},
  isbn       = {978-1-55860-737-8 978-0-08-050354-7 978-1-4933-0362-5},
  langid     = {american}
}

@misc{foundertypefoundryFounderOuyangXun,
  title        = {Founder {{Ouyang Xun Regular Script Font}}},
  author       = {{Founder Type Foundry}},
  urldate      = {2026-02-28},
  howpublished = {https://www.foundertype.com/index.php/FontInfo/index/id/6583.html},
  note         = {Accessed: 2026-02-28}
}

@inproceedings{fransCLIPDrawExploringTexttodrawing2022,
  title     = {{{CLIPDraw}}: {{Exploring}} Text-to-Drawing Synthesis through Language-Image Encoders},
  booktitle = {Advances in Neural Information Processing Systems},
  author    = {Frans, Kevin and Soros, Lisa and Witkowski, Olaf},
  editor    = {Koyejo, S. and Mohamed, S. and Agarwal, A. and Belgrave, D. and Cho, K. and Oh, A.},
  year      = 2022,
  volume    = {35},
  pages     = {5207--5218},
  publisher = {Curran Associates, Inc.},
  doi       = {10.52202/068431-0376}
}

@article{hirschornOptimizeReduceTopDown2024,
  title      = {Optimize \& {{Reduce}}: {{A Top-Down Approach}} for {{Image Vectorization}}},
  shorttitle = {Optimize \& {{Reduce}}},
  author     = {Hirschorn, Or and Jevnisek, Amir and Avidan, Shai},
  year       = 2024,
  month      = mar,
  journal    = {Proceedings of the AAAI Conference on Artificial Intelligence},
  volume     = {38},
  number     = {3},
  pages      = {2148--2156},
  issn       = {2374-3468},
  doi        = {10.1609/aaai.v38i3.27987},
  urldate    = {2026-02-20},
  copyright  = {Copyright (c) 2024 Association for the Advancement of Artificial Intelligence},
  langid     = {english}
}

@inproceedings{huang2DGaussianSplatting2024,
  title     = {{{2D Gaussian Splatting}} for {{Geometrically Accurate Radiance Fields}}},
  booktitle = {{{ACM SIGGRAPH}} 2024 {{Conference Papers}}},
  author    = {Huang, Binbin and Yu, Zehao and Chen, Anpei and Geiger, Andreas and Gao, Shenghua},
  year      = 2024,
  month     = jul,
  series    = {{{SIGGRAPH}} '24},
  pages     = {1--11},
  publisher = {Association for Computing Machinery},
  address   = {New York, NY, USA},
  doi       = {10.1145/3641519.3657428},
  urldate   = {2026-02-02},
  isbn      = {979-8-4007-0525-0}
}

@article{hughesIsogeometricAnalysisCAD2005,
  title      = {Isogeometric Analysis: {{CAD}}, Finite Elements, {{NURBS}}, Exact Geometry and Mesh Refinement},
  shorttitle = {Isogeometric Analysis},
  author     = {Hughes, T. J. R. and Cottrell, J. A. and Bazilevs, Y.},
  year       = 2005,
  month      = oct,
  journal    = {Computer Methods in Applied Mechanics and Engineering},
  volume     = {194},
  number     = {39},
  pages      = {4135--4195},
  issn       = {0045-7825},
  doi        = {10.1016/j.cma.2004.10.008},
  urldate    = {2026-02-26}
}

@article{iluzWordAsImageSemanticTypography2023,
  title   = {Word-{{As-Image}} for {{Semantic Typography}}},
  author  = {Iluz, Shir and Vinker, Yael and Hertz, Amir and Berio, Daniel and {Cohen-Or}, Daniel and Shamir, Ariel},
  year    = 2023,
  month   = jul,
  journal = {ACM Trans. Graph.},
  volume  = {42},
  number  = {4},
  pages   = {151:1--151:11},
  issn    = {0730-0301},
  doi     = {10.1145/3592123},
  urldate = {2026-02-26}
}

@inproceedings{jainVectorFusionTexttoSVGAbstracting2023,
  title      = {{{VectorFusion}}: {{Text-to-SVG}} by {{Abstracting Pixel-Based Diffusion Models}}},
  shorttitle = {{{VectorFusion}}},
  booktitle  = {2023 {{IEEE}}/{{CVF Conference}} on {{Computer Vision}} and {{Pattern Recognition}} ({{CVPR}})},
  author     = {Jain, Ajay and Xie, Amber and Abbeel, Pieter},
  year       = 2023,
  month      = jun,
  pages      = {1911--1920},
  issn       = {2575-7075},
  doi        = {10.1109/CVPR52729.2023.00190},
  urldate    = {2026-02-26}
}

@inproceedings{kaplanTSPArt2005,
  title     = {{{TSP Art}}},
  booktitle = {Renaissance {{Banff}}: {{Mathematics}}, {{Music}}, {{Art}}, {{Culture}}},
  author    = {Kaplan, Craig S. and Bosch, Robert},
  editor    = {Sarhangi, Reza and Moody, Robert V.},
  year      = 2005,
  pages     = {301--308},
  publisher = {Bridges Conference},
  address   = {Southwestern College, Winfield, Kansas},
  issn      = {1099-6702},
  isbn      = {0-9665201-6-5}
}

@misc{keikan&midnightgardenYoppaFudeFont,
  title        = {Yoppa {{Fude Font}}},
  author       = {{Keikan \& Midnight Garden}},
  urldate      = {2026-02-28},
  howpublished = {https://karu-k.booth.pm/items/2985842},
  note         = {Accessed: 2026-02-28}
}

@article{kerbl3DGaussianSplatting2023,
  title   = {{{3D Gaussian Splatting}} for {{Real-Time Radiance Field Rendering}}},
  author  = {Kerbl, Bernhard and Kopanas, Georgios and Leimkuehler, Thomas and Drettakis, George},
  year    = 2023,
  month   = aug,
  journal = {ACM Trans. Graph.},
  volume  = {42},
  number  = {4},
  pages   = {139:1--139:14},
  issn    = {0730-0301, 1557-7368},
  doi     = {10.1145/3592433},
  urldate = {2026-01-11},
  langid  = {english}
}

@inproceedings{kwonCLIPstylerImageStyle2022,
  title      = {{{CLIPstyler}}: {{Image Style Transfer}} with a {{Single Text Condition}}},
  shorttitle = {{{CLIPstyler}}},
  booktitle  = {2022 {{IEEE}}/{{CVF Conference}} on {{Computer Vision}} and {{Pattern Recognition}} ({{CVPR}})},
  author     = {Kwon, Gihyun and Ye, Jong Chul},
  year       = 2022,
  month      = jun,
  pages      = {18041--18050},
  issn       = {2575-7075},
  doi        = {10.1109/CVPR52688.2022.01753},
  urldate    = {2026-02-02}
}

@misc{liaoCalliffusionChineseCalligraphy2023,
  title         = {Calliffusion: {{Chinese Calligraphy Generation}} and {{Style Transfer}} with {{Diffusion Modeling}}},
  shorttitle    = {Calliffusion},
  author        = {Liao, Qisheng and Xia, Gus and Wang, Zhinuo},
  year          = 2023,
  month         = may,
  number        = {arXiv:2305.19124},
  eprint        = {2305.19124},
  primaryclass  = {cs},
  publisher     = {arXiv},
  doi           = {10.48550/arXiv.2305.19124},
  urldate       = {2026-02-23},
  archiveprefix = {arXiv}
}

@article{liDifferentiableVectorGraphics2020,
  title   = {Differentiable Vector Graphics Rasterization for Editing and Learning},
  author  = {Li, Tzu-Mao and Luk{\'a}{\v c}, Michal and Gharbi, Micha{\"e}l and {Ragan-Kelley}, Jonathan},
  year    = 2020,
  month   = dec,
  journal = {ACM Trans. Graph.},
  volume  = {39},
  number  = {6},
  pages   = {193:1--193:15},
  issn    = {0730-0301, 1557-7368},
  doi     = {10.1145/3414685.3417871},
  urldate = {2025-10-22},
  langid  = {english}
}

@inproceedings{liu2025bzier,
  title     = {B\'ezier Splatting for Fast and Differentiable Vector Graphics Rendering},
  booktitle = {Advances in Neural Information Processing Systems},
  author    = {Liu, Xi and Zhou, Chaoyi and Zhao, Nanxuan and Huang, Siyu},
  editor    = {Belgrave, D. and Zhang, C. and Lin, H. and Pascanu, R. and Koniusz, P. and Ghassemi, M. and Chen, N.},
  pages     = {51528--51559},
  publisher = {Curran Associates, Inc.},
  volume    = {38},
  year      = {2025},
  doi       = {10.52202/085713-1717}
}

@article{loopResolutionIndependentCurve2005,
  title   = {Resolution Independent Curve Rendering Using Programmable Graphics Hardware},
  author  = {Loop, Charles and Blinn, Jim},
  year    = 2005,
  month   = jul,
  journal = {ACM Trans. Graph.},
  volume  = {24},
  number  = {3},
  pages   = {1000--1009},
  issn    = {0730-0301},
  doi     = {10.1145/1073204.1073303},
  urldate = {2026-05-03}
}

@article{maAutomaticGenerationChinese1995,
  title   = {The Automatic Generation of Chinese Outline Font Based on Stroke Extraction},
  author  = {Ma, Xiaohu and Pan, Zhigeng and Zhang, Fuyan},
  year    = 1995,
  journal = {Journal of Computer Science and Technology},
  doi     = {10.1007/BF02939521},
  volume  = {10},
  number  = {1},
  pages   = {42--52},
  issn    = {1000-9000(Print) /1860-4749(Online)}
}

@inproceedings{maBezierGSDynamicUrban2025,
  title      = {{{B\'ezierGS}}: {{Dynamic Urban Scene Reconstruction}} with {{B\'ezier Curve Gaussian Splatting}}},
  shorttitle = {{{B\'ezierGS}}},
  booktitle  = {2025 {{IEEE}}/{{CVF International Conference}} on {{Computer Vision}} ({{ICCV}})},
  author     = {Ma, Zipei and Jiang, Junzhe and Chen, Yurui and Zhang, Li},
  year       = 2025,
  month      = oct,
  pages      = {25519--25528},
  issn       = {2380-7504},
  doi        = {10.1109/ICCV51701.2025.02367},
  urldate    = {2026-06-20}
}

@article{magneSingleLineDrawingVectorization2025,
  title     = {Single-{{Line Drawing Vectorization}}},
  author    = {Magne, Tanguy and {Sorkine-Hornung}, Olga},
  year      = 2025,
  journal   = {Comput. Graph. Forum},
  volume    = {44},
  number    = {7},
  pages     = {e70228},
  issn      = {1467-8659},
  doi       = {10.1111/cgf.70228},
  urldate   = {2026-01-31},
  copyright = {\copyright{} 2025 The Author(s). Computer Graphics Forum published by Eurographics - The European Association for Computer Graphics and John Wiley \& Sons Ltd.},
  langid    = {english}
}

@inproceedings{maLayerwiseImageVectorization2022,
  title     = {Towards {{Layer-wise Image Vectorization}}},
  booktitle = {2022 {{IEEE}}/{{CVF Conference}} on {{Computer Vision}} and {{Pattern Recognition}} ({{CVPR}})},
  author    = {Ma, Xu and Zhou, Yuqian and Xu, Xingqian and Sun, Bin and Filev, Valerii and Orlov, Nikita and Fu, Yun and Shi, Humphrey},
  year      = 2022,
  month     = jun,
  pages     = {16293--16302},
  publisher = {IEEE},
  address   = {New Orleans, LA, USA},
  doi       = {10.1109/CVPR52688.2022.01583},
  urldate   = {2026-02-02},
  copyright = {https://doi.org/10.15223/policy-029},
  isbn      = {978-1-6654-6946-3},
  langid    = {english}
}

@misc{ministryofeducationofthepeoplesrepublicofchinaTableGeneralStandard2013,
  title   = {{Table of General Standard Chinese Characters}},
  author  = {{Ministry of Education of the People's Republic of China}},
  year    = 2013,
  urldate = {2026-02-24},
  langid  = {chinese}
}

@misc{NotoEmoji2026,
  title        = {Noto {{Emoji}}},
  author       = {{Google Fonts}},
  urldate      = {2026-03-05},
  copyright    = {OFL-1.1},
  howpublished = {https://github.com/googlefonts/noto-emoji},
  note         = {Accessed: 2026-03-05}
}

@article{pottmannSmoothCurvesTension1990,
  title   = {Smooth Curves under Tension},
  author  = {Pottmann, H.},
  year    = 1990,
  month   = may,
  journal = {Computer-Aided Design},
  volume  = {22},
  number  = {4},
  pages   = {241--245},
  issn    = {0010-4485},
  doi     = {10.1016/0010-4485(90)90053-F},
  urldate = {2026-02-05}
}

@inproceedings{thamizharasanVecFusionVectorFont2024,
  title      = {{{VecFusion}}: {{Vector Font Generation}} with {{Diffusion}}},
  shorttitle = {{{VecFusion}}},
  booktitle  = {2024 {{IEEE}}/{{CVF Conference}} on {{Computer Vision}} and {{Pattern Recognition}} ({{CVPR}})},
  author     = {Thamizharasan, Vikas and Liu, Difan and Agarwal, Shantanu and Fisher, Matthew and Gharbi, Micha{\"e}l and Wang, Oliver and Jacobson, Alec and Kalogerakis, Evangelos},
  year       = 2024,
  month      = jun,
  pages      = {7943--7952},
  issn       = {2575-7075},
  doi        = {10.1109/CVPR52733.2024.00759},
  urldate    = {2026-02-23}
}

@article{tianSketchRefinerTextGuidedSketch2025,
  title      = {{{SketchRefiner}}: {{Text-Guided Sketch Refinement Through Latent Diffusion Models}}},
  shorttitle = {{{SketchRefiner}}},
  author     = {Tian, Yingjie and Liu, Minghao and Jiang, Haoran and Tu, Yunbin and Su, Duo},
  year       = 2025,
  month      = dec,
  journal    = {IEEE Transactions on Visualization and Computer Graphics},
  volume     = {31},
  number     = {12},
  pages      = {10711--10722},
  issn       = {1941-0506},
  doi        = {10.1109/TVCG.2025.3613388},
  urldate    = {2026-01-31}
}

@inproceedings{tojoFabricable3DWire2024,
  title     = {Fabricable {{3D Wire Art}}},
  booktitle = {{{ACM SIGGRAPH}} 2024 {{Conference Papers}}},
  author    = {Tojo, Kenji and Shamir, Ariel and Bickel, Bernd and Umetani, Nobuyuki},
  year      = 2024,
  month     = jul,
  series    = {{{SIGGRAPH}} '24},
  pages     = {1--11},
  publisher = {Association for Computing Machinery},
  address   = {New York, NY, USA},
  doi       = {10.1145/3641519.3657453},
  urldate   = {2026-02-26},
  isbn      = {979-8-4007-0525-0}
}

@article{vinkerCLIPassoSemanticallyawareObject2022,
  title      = {{{CLIPasso}}: Semantically-Aware Object Sketching},
  shorttitle = {{{CLIPasso}}},
  author     = {Vinker, Yael and Pajouheshgar, Ehsan and Bo, Jessica Y. and Bachmann, Roman Christian and Bermano, Amit Haim and {Cohen-Or}, Daniel and Zamir, Amir and Shamir, Ariel},
  year       = 2022,
  month      = jul,
  journal    = {ACM Trans. Graph.},
  volume     = {41},
  number     = {4},
  pages      = {86:1--86:11},
  issn       = {0730-0301},
  doi        = {10.1145/3528223.3530068},
  urldate    = {2026-02-26}
}

@article{wangDeepVecFontSynthesizingHighquality2021,
  title      = {{{DeepVecFont}}: Synthesizing High-Quality Vector Fonts via Dual-Modality Learning},
  shorttitle = {{{DeepVecFont}}},
  author     = {Wang, Yizhi and Lian, Zhouhui},
  year       = 2021,
  month      = dec,
  journal    = {ACM Trans. Graph.},
  volume     = {40},
  number     = {6},
  pages      = {265:1--265:15},
  issn       = {0730-0301},
  doi        = {10.1145/3478513.3480488},
  urldate    = {2026-02-23}
}

@inproceedings{wangDeepVecFontv2ExploitingTransformers2023,
  title      = {{{DeepVecFont-v2}}: {{Exploiting Transformers}} to {{Synthesize Vector Fonts}} with {{Higher Quality}}},
  shorttitle = {{{DeepVecFont-v2}}},
  booktitle  = {2023 {{IEEE}}/{{CVF Conference}} on {{Computer Vision}} and {{Pattern Recognition}} ({{CVPR}})},
  author     = {Wang, Yuqing and Wang, Yizhi and Yu, Longhui and Zhu, Yuesheng and Lian, Zhouhui},
  year       = 2023,
  month      = jun,
  pages      = {18320--18328},
  issn       = {2575-7075},
  doi        = {10.1109/CVPR52729.2023.01757},
  urldate    = {2026-02-23}
}

@inproceedings{wangImageSpaceCollagePacking2025,
  title     = {Image-{{Space Collage}} and {{Packing}} with {{Differentiable Rendering}}},
  booktitle = {Proceedings of the {{Special Interest Group}} on {{Computer Graphics}} and {{Interactive Techniques Conference Conference Papers}}},
  author    = {Wang, Zhenyu and Lu, Min},
  year      = 2025,
  numpages  = {11},
  articleno = {172},
  publisher = {Association for Computing Machinery},
  address   = {New York, NY, USA},
  doi       = {10.1145/3721238.3730690},
  urldate   = {2025-10-22},
  isbn      = {979-8-4007-1540-2},
  langid    = {english}
}

@inproceedings{wangLayeredImageVectorization2025,
  title     = {Layered {{Image Vectorization}} via {{Semantic Simplification}}},
  booktitle = {2025 {{IEEE}}/{{CVF Conference}} on {{Computer Vision}} and {{Pattern Recognition}} ({{CVPR}})},
  author    = {Wang, Zhenyu and Huang, Jianxi and Sun, Zhida and Gong, Yuanhao and {Cohen-Or}, Daniel and Lu, Min},
  year      = 2025,
  month     = jun,
  pages     = {7728--7738},
  issn      = {2575-7075},
  doi       = {10.1109/CVPR52734.2025.00724},
  urldate   = {2026-02-20}
}

@article{worchelDifferentiableRenderingParametric2023,
  title   = {Differentiable {{Rendering}} of {{Parametric Geometry}}},
  author  = {Worchel, Markus and Alexa, Marc},
  year    = 2023,
  month   = dec,
  journal = {ACM Trans. Graph.},
  volume  = {42},
  number  = {6},
  pages   = {232:1--232:18},
  issn    = {0730-0301},
  doi     = {10.1145/3618387},
  urldate = {2026-02-26}
}

@misc{wuCalliGANStyleStructureaware2020,
  title         = {{{CalliGAN}}: {{Style}} and {{Structure-aware Chinese Calligraphy Character Generator}}},
  shorttitle    = {{{CalliGAN}}},
  author        = {Wu, Shan-Jean and Yang, Chih-Yuan and Hsu, Jane Yung-jen},
  year          = 2020,
  month         = may,
  number        = {arXiv:2005.12500},
  eprint        = {2005.12500},
  primaryclass  = {cs},
  publisher     = {arXiv},
  doi           = {10.48550/arXiv.2005.12500},
  urldate       = {2026-02-23},
  archiveprefix = {arXiv}
}

@inproceedings{xingDiffSketcherTextGuided2023,
  title      = {{{DiffSketcher}}: {{Text Guided Vector Sketch Synthesis}} through {{Latent Diffusion Models}}},
  shorttitle = {{{DiffSketcher}}},
  booktitle  = {Advances in Neural Information Processing Systems},
  pages      = {15869--15889},
  publisher  = {Curran Associates, Inc.},
  author     = {Xing, XiMing and Wang, Chuang and Zhou, Haitao and Zhang, Jing and Yu, Qian and Xu, Dong},
  editor     = {Oh, A. and Naumann, T. and Globerson, A. and Saenko, K. and Hardt, M. and Levine, S.},
  volume     = {36},
  year       = {2023},
  doi        = {10.52202/075280-0698}
}

@inproceedings{xingSVGDreamerTextGuided2024,
  title      = {{{SVGDreamer}}: {{Text Guided SVG Generation}} with {{Diffusion Model}}},
  shorttitle = {{{SVGDreamer}}},
  booktitle  = {2024 {{IEEE}}/{{CVF Conference}} on {{Computer Vision}} and {{Pattern Recognition}} ({{CVPR}})},
  author     = {Xing, Ximing and Zhou, Haitao and Wang, Chuang and Zhang, Jing and Xu, Dong and Yu, Qian},
  year       = 2024,
  month      = jun,
  pages      = {4546--4555},
  issn       = {2575-7075},
  doi        = {10.1109/CVPR52733.2024.00435},
  urldate    = {2026-02-26}
}

@article{xuBrepGenBrepGenerative2024,
  title      = {{{BrepGen}}: {{A B-rep Generative Diffusion Model}} with {{Structured Latent Geometry}}},
  shorttitle = {{{BrepGen}}},
  author     = {Xu, Xiang and Lambourne, Joseph and Jayaraman, Pradeep and Wang, Zhengqing and Willis, Karl and Furukawa, Yasutaka},
  year       = 2024,
  month      = jul,
  journal    = {ACM Trans. Graph.},
  volume     = {43},
  number     = {4},
  pages      = {119:1--119:14},
  issn       = {0730-0301},
  doi        = {10.1145/3658129},
  urldate    = {2026-02-26}
}

@misc{yeIPAdapterTextCompatible2023,
  title         = {{{IP-Adapter}}: {{Text Compatible Image Prompt Adapter}} for {{Text-to-Image Diffusion Models}}},
  shorttitle    = {{{IP-Adapter}}},
  author        = {Ye, Hu and Zhang, Jun and Liu, Sibo and Han, Xiao and Yang, Wei},
  year          = 2023,
  month         = aug,
  number        = {arXiv:2308.06721},
  eprint        = {2308.06721},
  primaryclass  = {cs},
  publisher     = {arXiv},
  doi           = {10.48550/arXiv.2308.06721},
  urldate       = {2026-02-12},
  archiveprefix = {arXiv}
}

@inproceedings{yoonSplineGSLearningSmooth2024,
  title      = {{{SplineGS}}: {{Learning Smooth Trajectories}} in {{Gaussian Splatting}} for {{Dynamic Scene Reconstruction}}},
  shorttitle = {{{SplineGS}}},
  booktitle  = {The {{Thirteenth International Conference}} on {{Learning Representations}}},
  author     = {Yoon, Jihwan and Han, Sangbeom and Oh, Jaeseok and Lee, Minsik},
  year       = 2025,
  url        = {https://openreview.net/forum?id=tMG6btjBfd},
  note       = {Accessed: 2026-06-26}
}

@article{zengStrokeGANReducingMode2021,
  title      = {{{StrokeGAN}}: {{Reducing Mode Collapse}} in {{Chinese Font Generation}} via {{Stroke Encoding}}},
  shorttitle = {{{StrokeGAN}}},
  author     = {Zeng, Jinshan and Chen, Qi and Liu, Yunxin and Wang, Mingwen and Yao, Yuan},
  year       = 2021,
  month      = may,
  journal    = {Proceedings of the AAAI Conference on Artificial Intelligence},
  volume     = {35},
  number     = {4},
  pages      = {3270--3277},
  issn       = {2374-3468},
  doi        = {10.1609/aaai.v35i4.16438},
  urldate    = {2026-02-23},
  copyright  = {Copyright (c) 2021 Association for the Advancement of Artificial Intelligence},
  langid     = {english}
}

@inproceedings{zhangAddingConditionalControl2023,
  title     = {Adding {{Conditional Control}} to {{Text-to-Image Diffusion Models}}},
  booktitle = {2023 {{IEEE}}/{{CVF International Conference}} on {{Computer Vision}} ({{ICCV}})},
  author    = {Zhang, Lvmin and Rao, Anyi and Agrawala, Maneesh},
  year      = 2023,
  month     = oct,
  pages     = {3813--3824},
  issn      = {2380-7504},
  doi       = {10.1109/ICCV51070.2023.00355},
  urldate   = {2026-02-12}
}

@inproceedings{zhangGaussianImage1000FPS2025,
  title      = {{{GaussianImage}}: 1000 {{FPS Image Representation}} and~{{Compression}} by~{{2D Gaussian Splatting}}},
  shorttitle = {{{GaussianImage}}},
  booktitle  = {Computer {{Vision}} -- {{ECCV}} 2024},
  author     = {Zhang, Xinjie and Ge, Xingtong and Xu, Tongda and He, Dailan and Wang, Yan and Qin, Hongwei and Lu, Guo and Geng, Jing and Zhang, Jun},
  year       = 2025,
  editor     = {Leonardis, Ale{\v{s}}
                and Ricci, Elisa
                and Roth, Stefan
                and Russakovsky, Olga
                and Sattler, Torsten
                and Varol, G{\"u}l},
  pages      = {327--345},
  publisher  = {Springer Nature Switzerland},
  address    = {Cham},
  doi        = {10.1007/978-3-031-72673-6_18},
  isbn       = {978-3-031-72673-6},
  langid     = {english}
}

@inproceedings{zhangSketchDancingTextDrivenFramework2025,
  title      = {{{SketchDancing}}: {{A Text-Driven Framework}} for {{Vector Sketch Animation Generation}}},
  shorttitle = {{{SketchDancing}}},
  booktitle  = {Proceedings of the 3rd {{International Workshop}} on {{Multimodal}} and {{Responsible Affective Computing}}},
  author     = {Zhang, Xianlin and Liu, Zhuoyun and Liu, Jin and Li, Xueming and Qi, Mengshi},
  year       = 2025,
  month      = oct,
  series     = {{{MRAC}} '25},
  pages      = {128--136},
  publisher  = {Association for Computing Machinery},
  address    = {New York, NY, USA},
  doi        = {10.1145/3746270.3760235},
  urldate    = {2026-01-31},
  isbn       = {979-8-4007-2052-9}
}

@inproceedings{zhangUnreasonableEffectivenessDeep2018,
  title     = {The {{Unreasonable Effectiveness}} of {{Deep Features}} as a {{Perceptual Metric}}},
  booktitle = {2018 {{IEEE}}/{{CVF Conference}} on {{Computer Vision}} and {{Pattern Recognition}}},
  author    = {Zhang, Richard and Isola, Phillip and Efros, Alexei A. and Shechtman, Eli and Wang, Oliver},
  year      = 2018,
  month     = jun,
  pages     = {586--595},
  issn      = {2575-7075},
  doi       = {10.1109/CVPR.2018.00068},
  urldate   = {2026-02-26}
}

@article{zouSplineGenApproximatingUnorganized2025,
  title      = {{{SplineGen}}: {{Approximating}} Unorganized Points through Generative {{AI}}},
  shorttitle = {{{SplineGen}}},
  author     = {Zou, Qiang and Zhu, Lizhen and Wu, Jiayu and Yang, Zhijie},
  year       = 2025,
  month      = jan,
  journal    = {Computer-Aided Design},
  volume     = {178},
  pages      = {103809},
  issn       = {0010-4485},
  doi        = {10.1016/j.cad.2024.103809},
  urldate    = {2026-02-26}
}

\clearpage
\newgeometry{margin=1in}
\appendix

\setcounter{section}{0}
\renewcommand{\thesection}{S\arabic{section}}
\setcounter{equation}{0}
\renewcommand{\theequation}{S\arabic{equation}}
\setcounter{figure}{0}
\renewcommand{\thefigure}{S\arabic{figure}}
\setcounter{table}{0}
\renewcommand{\thetable}{S\arabic{table}}

\renewcommand{\theHsection}{supp.\arabic{section}}
\renewcommand{\theHsubsection}{supp.\arabic{section}.\arabic{subsection}}
\renewcommand{\theHsubsubsection}{supp.\arabic{section}.\arabic{subsection}.\arabic{subsubsection}}
\renewcommand{\theHfigure}{supp.\arabic{figure}}
\renewcommand{\theHtable}{supp.\arabic{table}}
\renewcommand{\theHequation}{supp.\arabic{equation}}

\begin{center}
  {\LARGE \textbf{Supplementary Materials}}
\end{center}

\section{Code and Data Availability}
Code will be made available upon publication at \url{https://github.com/AnicoderAndy/nurbs-splatting}.
All fonts used to construct our dataset are freely available online. However, due to licensing restrictions, some of these fonts cannot be redistributed directly. Instead, we will release the source code used for dataset construction.

\section{Signed Distance Field Evaluation}
The unsigned distance is the minimum Euclidean distance from $\mathbf{q}$ to the nearest boundary segment:
\begin{equation}
d(\mathbf{q}) = \min_{i} \left\| \mathbf{q} - \mathrm{proj}_{\overline{\mathbf{b}_i \mathbf{b}_{i+1}}}(\mathbf{q}) \right\|_2,
\end{equation}
where $\mathrm{proj}_{\overline{\mathbf{b}_i \mathbf{b}_{i+1}}}(\mathbf{q})$ denotes the closest point on segment $\overline{\mathbf{b}_i \mathbf{b}_{i+1}}$ to $\mathbf{q}$, computed by clamping the projection parameter to $[0,1]$.
The sign is determined by the winding number $\mathrm{wn}(\mathbf{q})$, which is detached from the computation graph to prevent gradient flow through the discrete sign:
\begin{gather}
\mathrm{wn}(\mathbf{q}) 
= \frac{1}{2\pi}
\sum_{i=1}^{M}
\angle
\left(
\mathbf{b}_i - \mathbf{q},\;
\mathbf{b}_{i+1} - \mathbf{q}
\right),\\
\mathrm{sdf}(\mathbf{q}) = \sgn\!\big(\mathrm{wn}(\mathbf{q})\big) \cdot d(\mathbf{q}),
\end{gather}
where $\sgn\left(\mathrm{wn}\right) = +1$ when $\mathrm{wn} > 0.5$ (inside) and $-1$ otherwise.
This formulation yields positive values inside the region and negative values outside, and the smooth unsigned distance provides well-behaved gradients near the boundary.

\section{Symbols and Settings}
We utilize numerous mathematical symbols throughout the main paper and supplementary material. For clarity, we summarize their definitions and default settings in \cref{tab:symbols}. In this table, ``Calligraphy Reconstruction'' is abbreviated as ``Cal.'', and ``Layer-wise Image Vectorization'' is abbreviated as ``LIVE''.

\begin{longtable}{c l l}
  \caption{Symbol definitions and default settings}
  \label{tab:symbols} \\
  \toprule
  Symbol & Description & Default \\
  \midrule
  \endfirsthead
  \toprule
  Symbol & Description & Default \\
  \midrule
  \endhead
  \bottomrule
  \endfoot
  \addlinespace
  \multicolumn{3}{l}{\textbf{NURBS Curve Parameters}}\\
  $p$ & Degree of NURBS curve & 5 (Cal.), 3 (LIVE) \\
  $k{=}p{+}1$ & Order of NURBS curve & 6 (Cal.), 4 (LIVE) \\
  $n{+}1$ & Number of control points & Per-curve \\
  $n_k$ & Number of key points & Per-curve; 12 (LIVE init) \\
  $\mathbf{P}_i$ & Control points $\in\mathbb{R}^2$ & Learnable \\
  $w_i$ & Rational weight per control point & Init.\ 1.0; clamped to $[0.01, 10]$ \\
  $\tau_j$ & Learnable knot interval & Init.\ uniform; clamped to $[0, 2]$ \\
  $\mathbf{U}$ & Knot vector & Derived from $\{\tau_j\}$ \\
  $\mathbf{c}$ & Curve RGB color $\in[0,1]^3$ & $[0,0,0]$ (Cal.) \\
  $o$ & Curve opacity $\in[0,1]$ & 1.0 \\
  \addlinespace
  \multicolumn{3}{l}{\textbf{Gaussian Contour Sampling}} \\
  $L$ & Arc length of the curve & Computed \\
  $N_L$ & Samples for arc-length quadrature & $3\times\max(W,H)$\footnote{$W,H$ are the width and height of the image.} \\
  $M$ & Number of contour Gaussians & $\lceil \delta_\text{c} \cdot L\rceil$ \\
  $\delta_\text{c}$ & Contour density\footnote{$\delta_\text{c}=10$ is sufficient to capture curve details and reduce blurry edges in our tests. However, we use $\delta_\text{c}=18$ for better quality in calligraphy reconstruction.} (samples per px arc length) & 10 \\
  $\rho$ & Global Gaussian scale ratio & 2.0 \\
  $\sigma_j$ & Isotropic Gaussian std.\ dev. & $s_j / \rho$ \\
  $s_j$ & Local stroke width at parameter $u_j$ & Learnable per control point \\
  $\kappa$ & Replication count for endpoint Gaussians & 4 \\
  \addlinespace
  \multicolumn{3}{l}{\textbf{Interior Fill (Closed Curves)}} \\
  $\delta_\text{b}$ & Boundary sample density & $0.5$ \\
  $h$ & Fill grid step (pixels) & 1.0 \\
  $h_{\max}$, $h_{\min}$ & Grid-step annealing bounds & 4.0, 1.0 (LIVE) \\
  $t_b$, $t_e$ & Annealing begin/end fractions & 0.1, 0.8 (LIVE) \\
  $\eta$ & Coverage factor for fill Gaussians & 1.5 \\
  $\gamma$ & SDF sigmoid sharpness & $5.0$ \\
  $Q{=}n_x{\times}n_y$ & Number of interior grid points & Computed from $h$ \\
  \addlinespace
  \multicolumn{3}{l}{\textbf{Differentiable Splatting}} \\
  $\boldsymbol{\Sigma}_i$ & Gaussian covariance matrix & $\sigma_i^2 \mathbf{I}$ (isotropic) \\
  $\alpha_i(\mathbf{x}_n)$ & Per-pixel Gaussian opacity & Eq.~9 in Sec.~3.4\\
  $\mathcal{I}(\mathbf{x}_n)$ & Rendered image at pixel $\mathbf{x}_n$ & --- \\
  $\mathcal{I}^{\ast}$ & Target image & --- \\
  \addlinespace
  \multicolumn{3}{l}{\textbf{SDF Evaluation (Supplementary)}} \\
  $\mathbf{q}$ & Query point for SDF evaluation & Grid point \\
  $\mathbf{b}_i$ & Boundary sample points & From curve sampling \\
  $d(\mathbf{q})$ & Unsigned distance to boundary & Computed \\
  $\mathrm{wn}(\mathbf{q})$ & Winding number & Computed; detached \\
  $\mathrm{sdf}(\mathbf{q})$ & Signed distance field value & $\sgn(\mathrm{wn}) \cdot d(\mathbf{q})$ \\
  \addlinespace
  \multicolumn{3}{l}{\textbf{Loss Functions and Optimization}} \\
  $r$ & Derivative order for smoothing energy & 3 \\
  $\mathcal{L}_{\text{deriv}}^r$ & $r$-th order derivative loss & --- \\
  $\mathcal{L}_{\text{bbox}}$ & Bounding-box penalty & --- \\
  $\mathcal{L}_{\text{xing}}$ & Self-crossing penalty & --- \\
  $\lambda_{\text{MSE}}$ & MSE loss weight & 1.0 (Cal.) \\
  $\lambda_{\text{deriv}}$ & Derivative loss weight & 1.0 (Cal.) \\
  $\lambda_{\text{bbox}}$ & Bounding-box loss weight & 5.0 (Cal.) \\
  $\lambda_{\text{xing}}$ & Crossing loss weight & 0.01 (LIVE) \\
\end{longtable}

\section{Experiments}
We provide additional details on the efficiency test (\cref{sec:efficiency-test}), the editability demonstration (\cref{sec:editability}), the experimental setup and results for our calligraphy reconstruction experiment~(\cref{sec:calligraphy-details}) and the LIVE experiment~(\cref{sec:live-details}), as well as further discussion on our comparison with B\'{e}zier Splatting~(\cref{sec:no-bs-comparison}).

\subsection{Efficiency Test}
\label{sec:efficiency-test}

\begin{figure}[t]
    \centering
    \footnotesize

    \begin{minipage}[t]{0.33\linewidth}
        \vspace{20pt}
        \centering
        \begin{tabular}{c cc cc}
            \toprule
            \multirow{2}{*}[-0.3em]{Res.}
                & \multicolumn{2}{c}{Forward}
                & \multicolumn{2}{c}{Backward} \\
            \cmidrule(lr){2-3}
            \cmidrule(lr){4-5}
                & Ours & B. & Ours & B. \\
            \midrule
            128  & \textbf{7.136} & 1452 & \textbf{2.495} & 18.20 \\
            512  & \textbf{7.020} & 1488 & \textbf{3.313} & 100.6 \\
            1024 & \textbf{6.943} & 1426 & \textbf{3.657} & 223.2 \\
            4096 & \textbf{10.67} & 1574 & \textbf{8.767} & 2538 \\
            \bottomrule
        \end{tabular}
    \end{minipage}
    \hfill
    \begin{minipage}[t]{0.31\linewidth}
        \vspace{0pt}
        \centering
        \includegraphics[width=\linewidth]{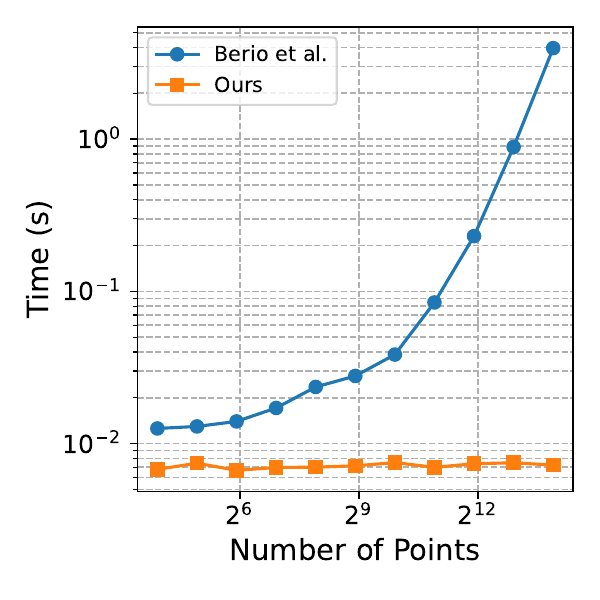}
    \end{minipage}
    \hfill
    \begin{minipage}[t]{0.31\linewidth}
        \vspace{0pt}
        \centering
        \includegraphics[width=\linewidth]{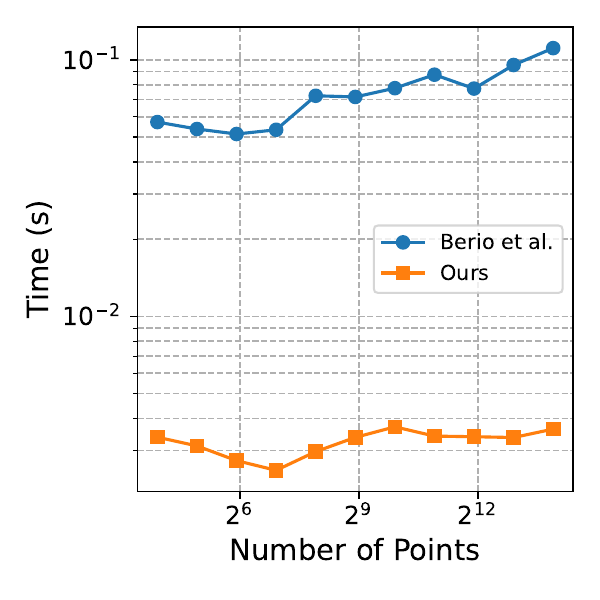}
    \end{minipage}

    \caption{
    Runtime analysis. The table reports the forward and backward execution time (ms) for optimizing a single curve with 10,240 key points under different rendering resolutions. The two plots show the forward (middle) and backward (right) execution time at $512\times512$ while varying the number of control points ($N$).
    }
    \label{fig:scalability}
\end{figure}

\begin{table}[t]
  \centering
  \footnotesize
  \caption{Speed comparison of forward and backward passes, as well as total time (in ms), under different rendering resolutions. All methods render 16 curves. For a fair comparison with the official implementation of B\'ezier Splatting, all curves are cubic (degree 3); open curves use 10 control points, whereas closed curves use 6. \cellcolor{red!25}Best results are highlighted in \colorbox{red!25}{red} and second-best in \colorbox{yellow!25}{yellow}.}
  \label{tab:speeds}
  \begin{tabular}{l ccc ccc ccc}
    \toprule
    & \multicolumn{3}{c}{Ours} & \multicolumn{3}{c}{Berio \etal~\cite{berioNeuralImageAbstraction2025}} & \multicolumn{3}{c}{B\'ezier Splatting~\cite{liu2025bzier}} \\
    \cmidrule(lr){2-4} \cmidrule(lr){5-7} \cmidrule(lr){8-10}
    Res. & Fwd & Bwd & Total & Fwd & Bwd & Total & Fwd & Bwd & Total \\
    \midrule
    \multicolumn{10}{c}{\textbf{Closed Curves}} \\
    \midrule
    128$\times$128 & 97.42 & 43.15 & 140.57 & \cellcolor{red!25}24.69 & \cellcolor{yellow!25}17.30 & \cellcolor{yellow!25}41.99 & \cellcolor{yellow!25}28.27 & \cellcolor{red!25}4.29 & \cellcolor{red!25}32.56 \\
    256$\times$256 & 96.35 & 38.17 & 134.52 & \cellcolor{red!25}26.05 & \cellcolor{yellow!25}25.42 & \cellcolor{yellow!25}51.47 & \cellcolor{yellow!25}27.81 & \cellcolor{red!25}5.61 & \cellcolor{red!25}33.42 \\
    512$\times$512 & 96.24 & \cellcolor{yellow!25}39.00 & 135.24 & \cellcolor{yellow!25}38.84 & 62.15 & \cellcolor{yellow!25}100.99 & \cellcolor{red!25}27.65 & \cellcolor{red!25}4.22 & \cellcolor{red!25}31.87 \\
    1024$\times$1024 & 95.73 & \cellcolor{yellow!25}43.52 & \cellcolor{yellow!25}139.25 & \cellcolor{yellow!25}45.13 & 204.49 & 249.62 & \cellcolor{red!25}27.93 & \cellcolor{red!25}5.56 & \cellcolor{red!25}33.49 \\
    \midrule
    \multicolumn{10}{c}{\textbf{Open Curves}} \\
    \midrule
    128$\times$128 & 72.69 & 32.57 & 105.26 & \cellcolor{yellow!25}25.23 & \cellcolor{yellow!25}19.33 & \cellcolor{yellow!25}44.56 & \cellcolor{red!25}4.53 & \cellcolor{red!25}3.64 & \cellcolor{red!25}8.17 \\
    256$\times$256 & 72.36 & 32.23 & 104.59 & \cellcolor{yellow!25}25.25 & \cellcolor{yellow!25}27.34 & \cellcolor{yellow!25}52.59 & \cellcolor{red!25}4.28 & \cellcolor{red!25}3.31 & \cellcolor{red!25}7.59 \\
    512$\times$512 & 74.00 & \cellcolor{yellow!25}32.33 & 106.33 & \cellcolor{yellow!25}40.79 & 60.38 & \cellcolor{yellow!25}101.17 & \cellcolor{red!25}4.19 & \cellcolor{red!25}3.22 & \cellcolor{red!25}7.41 \\
    1024$\times$1024 & 74.23 & \cellcolor{yellow!25}33.92 & \cellcolor{yellow!25}108.15 & \cellcolor{yellow!25}41.81 & 193.41 & 235.22 & \cellcolor{red!25}4.76 & \cellcolor{red!25}4.53 & \cellcolor{red!25}9.29 \\
    \bottomrule
  \end{tabular}
\end{table}

We conduct an efficiency test to evaluate the computational performance of our method compared to the baseline. \cref{tab:speeds} shows the comparison of forward/backward passes and the total time for each method under different resolution settings.
It is shown that our method is slightly slower than the baseline methods at lower resolutions.
Based on the profiling results, we hypothesize that the performance gap primarily stems from the evaluation of NURBS, which appears to dominate the overall runtime. Compared with methods based on B\'ezier curves or uniform B-splines, NURBS evaluation is inherently more computationally demanding because of its higher computational complexity, making it a likely performance bottleneck.
Additionally, B\'ezier Splatting exhibits superior scalability as the number of geometric primitives increases. This advantage may be attributed to its representation of all curves as a single fixed-shape tensor, which enables efficient batched GPU computation but requires all curves to share the same geometric structure, such as an identical number of control points.

Nevertheless, our method demonstrates promising scalability as the number of control points increases, outperforming Berio \etal~\cite{berioNeuralImageAbstraction2025} in terms of efficiency. As shown in Fig.~\ref{fig:scalability}, our method exhibits a more favorable runtime trend as the number of control points increases. We attribute this advantage to the fact that Berio \etal\ first convert B-spline segments into Bézier curves through Bézier extraction (implemented as a sequence of local matrix operations) before evaluation, introducing additional preprocessing overhead. In contrast, our method operates directly on the original NURBS representation, eliminating the need for such conversions. Consequently, although Bézier extraction has linear complexity with respect to the number of control points, its additional preprocessing cost accumulates as curve complexity increases. By avoiding this conversion entirely, our method scales more gracefully with the number of control points.

\subsection{Editability Demonstration}
\label{sec:editability}
\begin{figure}[t]
  \centering
  \setlength{\tabcolsep}{1pt}
  \begin{tabular}{cccc}
    \includegraphics[width=0.24\linewidth]{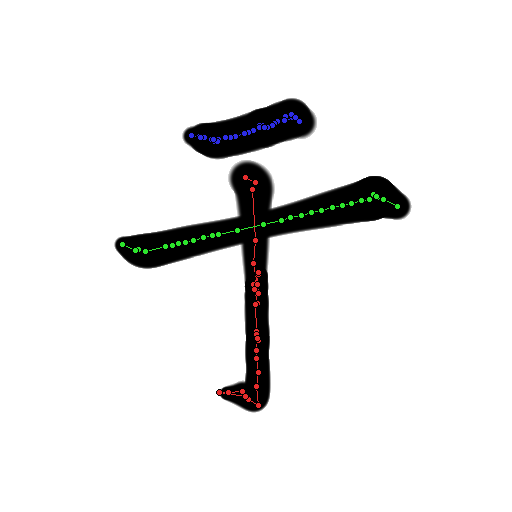} &
    \includegraphics[width=0.24\linewidth]{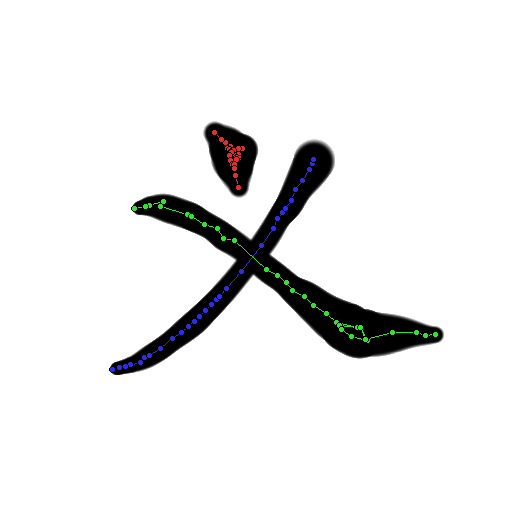} &
    \includegraphics[width=0.24\linewidth]{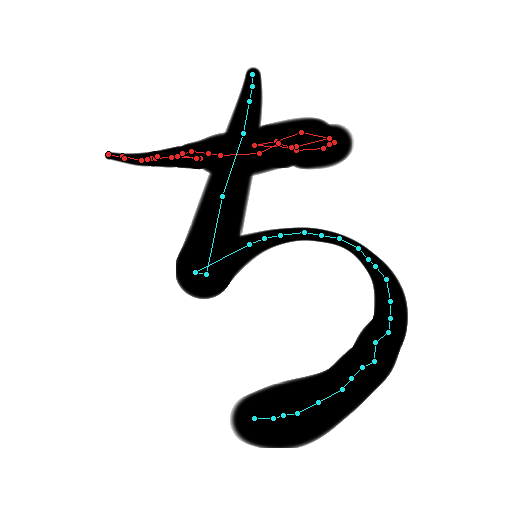} &
    \includegraphics[width=0.24\linewidth]{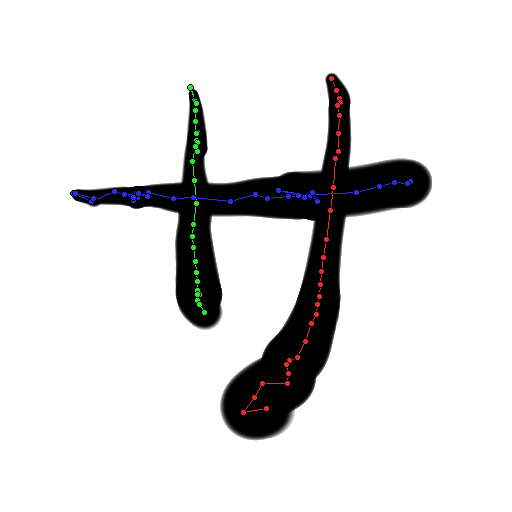} \\
  \end{tabular}
  \caption{Control points of NURBS curves optimized for calligraphy reconstruction.}
  \label{fig:editability}
\end{figure}

\cref{fig:editability} shows the control points of the NURBS curves optimized for the calligraphy reconstruction task. Since mainstream vector graphics editors generally do not support NURBS editing, providing an editing demonstration in existing software is impractical. Instead, we visualize the optimized control-point layout, which remains compact and interpretable for manual adjustment.

\subsection{Calligraphy Reconstruction}
\label{sec:calligraphy-details}
We extract initial curves from the input calligraphy images using SLDVec~\cite{magneSingleLineDrawingVectorization2025}. Because SLDVec generates a dense set of $n_\text{SLD}$ points per stroke, we downsample each stroke to $\max(30, 0.01 n_\text{SLD})$ control points to improve optimization efficiency.
The dataset consists of 192 characters, one of which failed the extraction process and was therefore excluded from all experiments.

Both our method and the Berio~\etal~\cite{berioNeuralImageAbstraction2025} baseline optimize degree $p{=}5$ curves for 150 Adam iterations. We apply cosine learning-rate annealing, decaying the rate to $10\%$ of its initial value. For our method, we set the learning rates to 1.5 for control-point positions, 0.2 for stroke widths, and 0.1 for both rational weights and knot intervals. We clamp the rational weights to $[0.01, 10]$ and knot intervals to $[0, 2]$. The baseline uses the same position learning rate but a slightly higher width learning rate (0.3), and it does not optimize rational weights or knots.

We weight our objective function components with $\lambda_\text{MSE}{=}1$, $\lambda_\text{deriv}{=}1$, and $\lambda_\text{bbox}{=}5$. The baseline uses the DiffVG rasterizer~\cite{liDifferentiableVectorGraphics2020} with independently tuned loss weights ($\lambda_\text{MSE}{=}20$, $\lambda_\text{deriv}{=}3$, $\lambda_\text{bbox}{=}10$).

We evaluate reconstruction quality using pixel-level and stroke-geometry metrics. To measure pixel-level fidelity, we compute MSE, PSNR, and SSIM between the rendered output and the ground-truth raster image. To evaluate stroke geometry, we binarize the rendered and target images at an intensity threshold of $0.5$ to extract stroke masks. We then calculate the Hausdorff distance between the edge pixels of these masks, as well as the F1 score.

\cref{tab:calligraphy-metrics-detailed} reports the detailed quantitative results for this application.

\begin{table}[t]
  \centering
  \setlength{\tabcolsep}{3pt}
  \caption{Quantitative comparison of calligraphy reconstruction quality and runtime. ``Ext.'' is the time for SLDVec~\cite{magneSingleLineDrawingVectorization2025} extraction. The baseline is Berio \etal~\cite{berioNeuralImageAbstraction2025}. ``Proto. w/o w \& k'' is our implemented prototype using the same contour-sampling strategy as B\'{e}zier Splatting~\cite{liu2025bzier}, which is mentioned in \cref{sec:no-bs-comparison}. \cellcolor{red!25}Best results are highlighted in \colorbox{red!25}{red} and second-best in \colorbox{yellow!25}{yellow}.}
  \begin{tabular}{l l c c c c c c c}
    \toprule
    & & \multicolumn{5}{c}{Quality Metrics} & \multicolumn{2}{c}{Runtime (s)} \\
    \cmidrule(lr){3-7} \cmidrule(lr){8-9}
    Dataset & Method & MSE$\downarrow$ & PSNR$\uparrow$ & SSIM$\uparrow$ & Hausdorff$\downarrow$ & F1$\uparrow$ & Ext. & Opt. \\
    \midrule
    \multirow{7}{*}{All} & Baseline & 0.0057 & 24.72 & \cellcolor{yellow!25}0.9813 & 15.26 & 0.9642 & \multirow{7}{*}{13.8} & 10.65 \\
     & Ours & \cellcolor{red!25}0.0038 & \cellcolor{red!25}26.32 & 0.9793 & \cellcolor{red!25}10.69 & \cellcolor{yellow!25}0.9741 & & 8.73 \\
     & Ours ($\delta_\text{c}{=}30$) & \cellcolor{yellow!25}0.0039 & \cellcolor{yellow!25}26.24 & \cellcolor{red!25}0.9824 & 11.07 & \cellcolor{red!25}0.9743 & & 9.34 \\
     & Ours w/o knot & 0.0040 & 26.07 & 0.9789 & \cellcolor{yellow!25}10.98 & 0.9731 & & 5.63 \\
     & Ours w/o weight & 0.0046 & 25.63 & 0.9777 & 11.70 & 0.9701 & & 8.98 \\
     & Ours w/o w \& k & 0.0042 & 25.92 & 0.9785 & 11.34 & 0.9721 & & \cellcolor{yellow!25}5.12 \\
     & Proto. w/o w \& k & 0.0168 & 18.75 & 0.9060 & 25.70 & 0.8914 & & \cellcolor{red!25}3.94 \\
    \midrule
    \multirow{7}{*}{Japanese} & Baseline & 0.0075 & 23.21 & \cellcolor{yellow!25}0.9765 & 17.66 & 0.9618 & \multirow{7}{*}{13.9} & 10.66 \\
     & Ours & \cellcolor{red!25}0.0052 & \cellcolor{red!25}25.31 & 0.9748 & \cellcolor{red!25}13.93 & \cellcolor{red!25}0.9712 & & 7.92 \\
     & Ours ($\delta_\text{c}{=}30$) & \cellcolor{yellow!25}0.0053 & \cellcolor{yellow!25}25.28 & \cellcolor{red!25}0.9786 & 14.47 & \cellcolor{yellow!25}0.9712 & & 8.56 \\
     & Ours w/o knot & 0.0055 & 24.93 & 0.9740 & \cellcolor{yellow!25}14.15 & 0.9695 & & 5.27 \\
     & Ours w/o weight & 0.0064 & 24.45 & 0.9726 & 15.45 & 0.9657 & & 8.49 \\
     & Ours w/o w \& k & 0.0057 & 24.84 & 0.9736 & 14.57 & 0.9685 & & \cellcolor{yellow!25}4.78 \\
     & Proto. w/o w \& k & 0.0226 & 17.18 & 0.8879 & 29.65 & 0.8697 & & \cellcolor{red!25}3.81 \\
    \midrule
    \multirow{7}{*}{Chinese} & Baseline & 0.0040 & 26.13 & \cellcolor{yellow!25}0.9859 & 13.03 & 0.9665 & \multirow{7}{*}{13.7} & 10.64 \\
     & Ours & \cellcolor{red!25}0.0026 & \cellcolor{red!25}27.26 & 0.9835 & \cellcolor{red!25}7.67 & \cellcolor{yellow!25}0.9768 & & 9.48 \\
     & Ours ($\delta_\text{c}{=}30$) & \cellcolor{yellow!25}0.0026 & \cellcolor{yellow!25}27.13 & \cellcolor{red!25}0.9859 & \cellcolor{yellow!25}7.92 & \cellcolor{red!25}0.9773 & & 10.06 \\
     & Ours w/o knot & 0.0026 & 27.13 & 0.9835 & 8.03 & 0.9764 & & 5.96 \\
     & Ours w/o weight & 0.0029 & 26.72 & 0.9825 & 8.21 & 0.9741 & & 9.44 \\
     & Ours w/o w \& k & 0.0027 & 26.92 & 0.9831 & 8.35 & 0.9755 & & \cellcolor{yellow!25}5.44 \\
     & Proto. w/o w \& k & 0.0113 & 20.21 & 0.9228 & 22.03 & 0.9115 & & \cellcolor{red!25}4.05 \\
    \bottomrule
  \end{tabular}
  \label{tab:calligraphy-metrics-detailed}
\end{table}

\subsection{Layer-wise Image Vectorization}
\label{sec:live-details}
In our Layer-wise Image Vectorization (LIVE) experiments, we replace the cubic B\'{e}zier paths from the original LIVE framework~\cite{maLayerwiseImageVectorization2022} with our closed, filled NURBS curves. For the LIVE baseline, we retain all of its default hyperparameter settings with the sole exception of the initial circle radius, which we set to 16 pixels. Our method also uses an initial circle radius of 16 pixels, with our specific hyperparameter settings and optimization schedule detailed below.

Each of our curves is initialized as a circle using 12 key points and a degree of $p{=}3$. For the optimization process, we train each path addition for 300 Adam iterations. We apply learning rate decay with a ratio of 0.4 over the course of the iterations. The initial learning rates are set to 2.0 for control-point positions, and 0.01 for colors, rational weights, knot intervals, and opacities. In addition to the standard reconstruction loss, our objective function incorporates a self-crossing penalty formulated in the original LIVE method with weight $\lambda_\text{xing}{=}0.01$. Note that the third-order derivative smoothing loss is not activated in this experiment because the curve degree is $p{=}3$.

To accelerate the differentiable filling process while preserving fine boundary details, we employ grid-step annealing. Within each 300-iteration round, the fill grid step $h$ is cosine-annealed from $h_{\max}{=}4.0$ to $h_{\min}{=}1.0$ between $10\%$ and $80\%$ of the iterations (\ie, $t_b{=}0.1$ and $t_e{=}0.8$).

\paragraph{Discussion on Performance.} Our method achieves both higher efficiency and better reconstruction quality than the LIVE baseline. Readers may notice an apparent inconsistency: the results in \cref{sec:efficiency-test} indicate that the Berio \etal baseline (based on DiffVG~\cite{liDifferentiableVectorGraphics2020}) is faster than our method, whereas in this experiment our approach outperforms the baseline (also based on DiffVG) in both speed and reconstruction quality.
We believe that two factors contribute to this difference. (a) We employ a customized \texttt{udf\_weight} loss computation tailored to our framework. Unlike the LIVE baseline, which evaluates the SDF on the CPU, our implementation performs this computation on the GPU, resulting in substantially higher efficiency. (b) Our step grid annealing strategy partially alleviates the excessive Gaussian generation caused by region filling, thereby improving both optimization efficiency and reconstruction performance.

\subsection{Discussion on B\'{e}zier Splatting}
\label{sec:no-bs-comparison}
B\'ezier Splatting~(BS)~\cite{liu2025bzier} demonstrates excellent efficiency for vector graphic reconstruction and, to the best of our knowledge, remains the current state-of-the-art method in terms of computational performance. Nevertheless, its representation is less flexible than ours in several aspects. First, it focuses on vectorization tasks and is temporarily not well-suited for other applications.
Second, its paired B\'ezier representation requires each filled region to consist of \emph{pairs} of curves. Boundaries with an odd number of segments (\eg, a triangle) therefore require an additional dummy point to complete the pairing, introducing an unnecessary degree of freedom that may adversely affect gradient-based optimization.
Third, as discussed in \cref{sec:efficiency-test}, BS requires all curves in a scene to share the same geometric configuration, such as the number of control points, whereas our method allows heterogeneous curve representations within a single scene.

We have included a quantitative comparison of computational performance between our method and BS in \cref{sec:efficiency-test}, as well as a qualitative comparison of rendering quality in Sec.~5.1 of the main paper. However, we did not include quantitative comparisons on downstream tasks. The reasons are explained below.

\smallskip
\noindent\emph{Calligraphy reconstruction.}
The publicly released BS codebase requires all B\'{e}zier curves in a scene to share a fixed number of control points.
Our calligraphy benchmark, however, contains strokes extracted via SLDVec~\cite{magneSingleLineDrawingVectorization2025} with varying numbers of control points per stroke, reflecting the inherent complexity differences among strokes within a single character.
This constraint makes it infeasible to run BS on our dataset without substantial modification of its optimization pipeline.
Moreover, BS operates on cubic B\'{e}zier curves only and does not support B-splines, varying stroke widths, or long splines---all of which are central to our calligraphy formulation.
These architectural differences preclude an apples-to-apples comparison under identical experimental conditions.

\begin{figure}[t]
  \centering
  \def\width{0.16\linewidth}
  \begin{tabular}{@{}c@{\hspace{1pt}}c@{\hspace{1pt}}c@{\hspace{4pt}}c@{\hspace{1pt}}c@{\hspace{1pt}}c@{}}
    \includegraphics[width=\width]{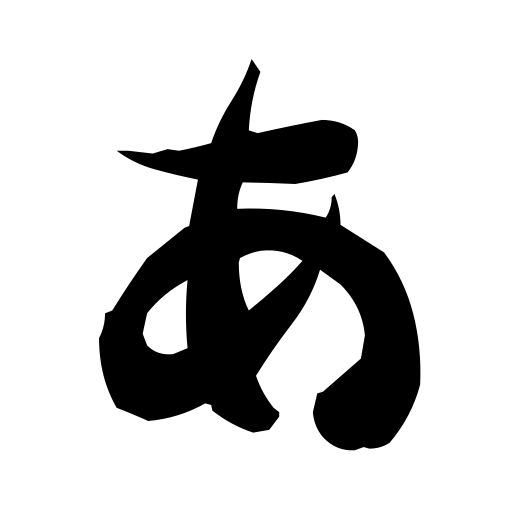} &
    \includegraphics[width=\width]{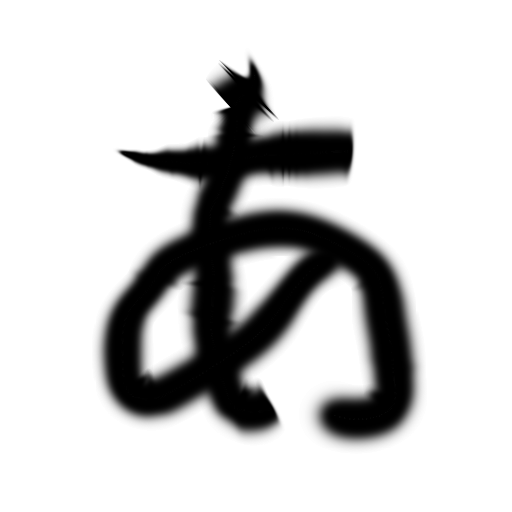} &
    \includegraphics[width=\width]{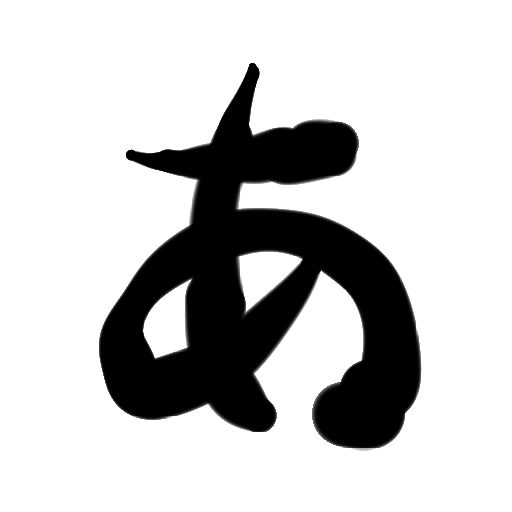} &
    \includegraphics[width=\width]{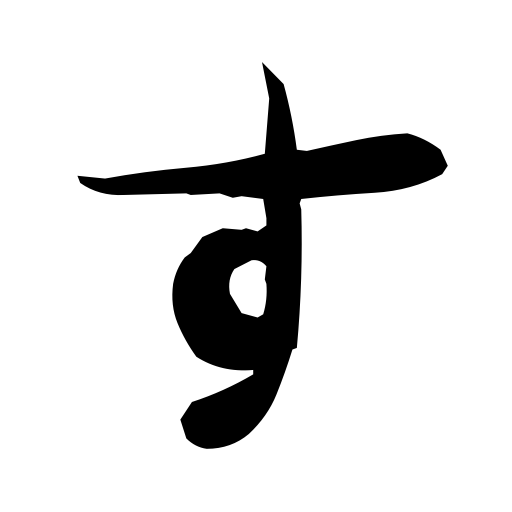} &
    \includegraphics[width=\width]{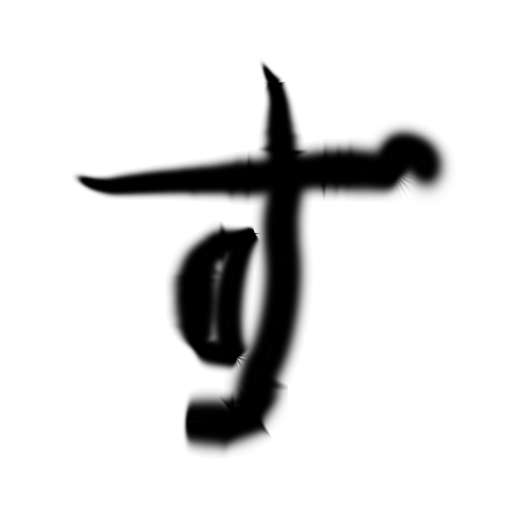} &
    \includegraphics[width=\width]{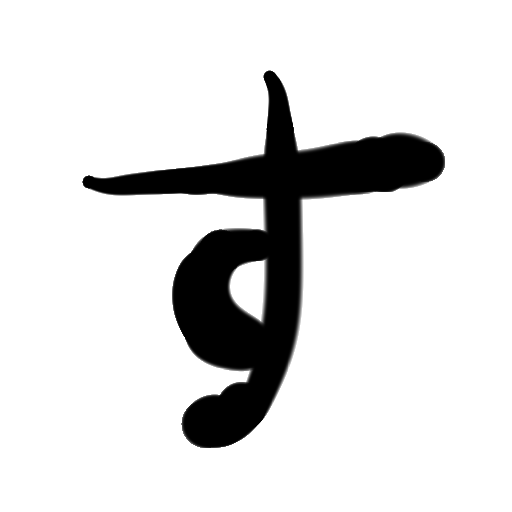} \\
    \includegraphics[width=\width]{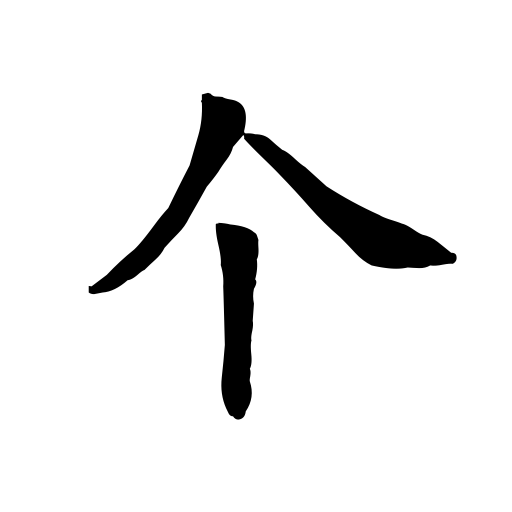} &
    \includegraphics[width=\width]{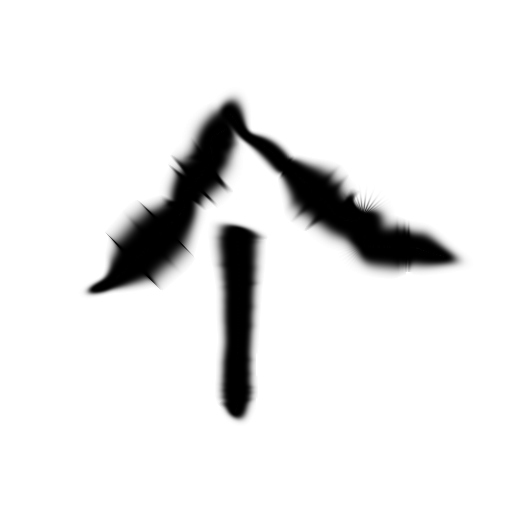} &
    \includegraphics[width=\width]{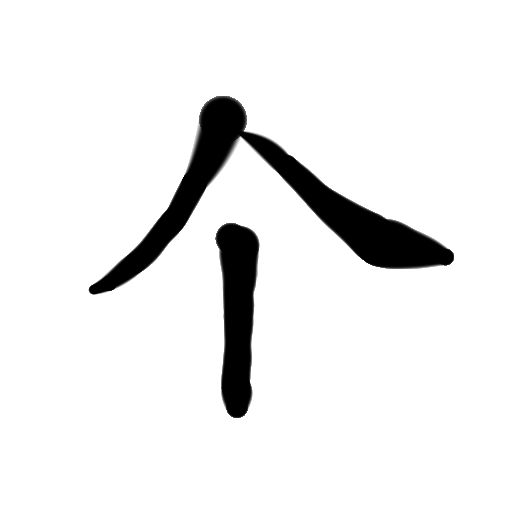} &
    \includegraphics[width=\width]{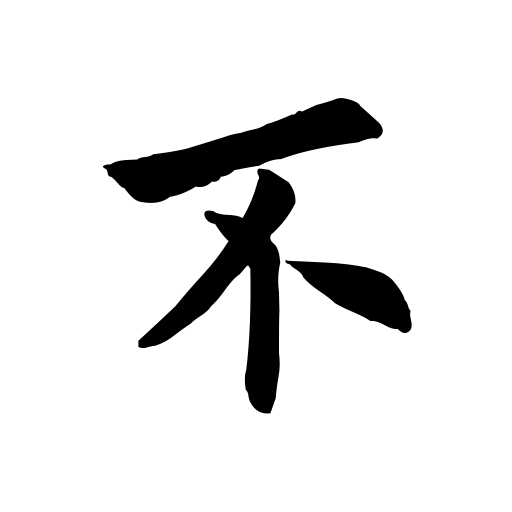} &
    \includegraphics[width=\width]{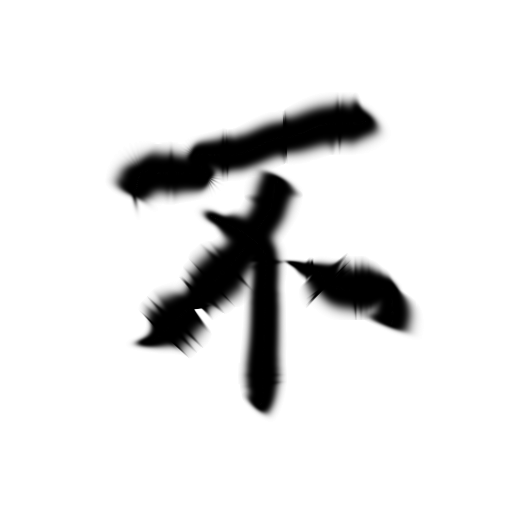} &
    \includegraphics[width=\width]{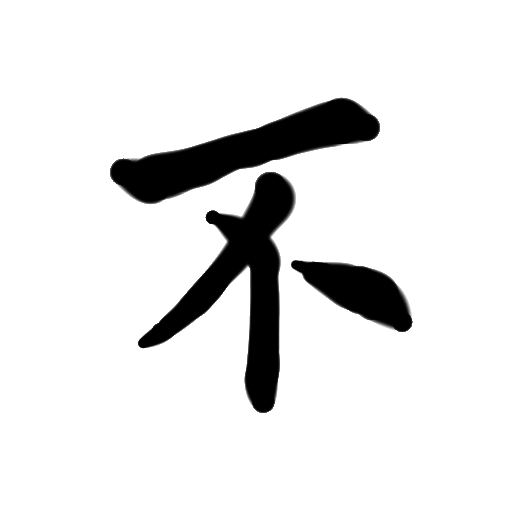} \\
    Target & BS Proto. & Ours & Target & BS Proto. & Ours
  \end{tabular}
  \caption{Qualitative comparison on calligraphy reconstruction against an early prototype utilizing the anisotropic Gaussian contour-sampling strategy of B\'{e}zier Splatting (BS Proto.). Both BS Proto.\ and Ours disable weights and knots optimization for this comparison. The top row displays Japanese samples, while the bottom row displays Chinese examples.}
  \label{fig:calligraphy_bs}
\end{figure}

Nonetheless, to provide a partial reference, we note that an early prototype of our framework employed the same contour-sampling strategy as BS (\ie, anisotropic Gaussians aligned to curve tangents).
We test this prototype on our calligraphy benchmark and the results are included in \cref{tab:calligraphy-metrics-detailed}. As shown in the table, while the anisotropic sampling strategy achieves slightly faster optimization due to fewer splats, it significantly degrades reconstruction quality, with lower PSNR and higher Hausdorff distances compared to our final formulation. Furthermore, as shown in \cref{fig:calligraphy_bs}, the BS prototype completely fails to preserve the intricate stroke boundaries and varying widths of the original calligraphy.

\smallskip
\noindent\emph{Layer-wise image vectorization.}
The BS authors have not released their code for the layer-wise vectorization application described in their paper. Consequently, no compatible implementation is available for us to evaluate on our LIVE benchmark.

\smallskip
\noindent\emph{Summary.}
Our qualitative comparison in the main paper, alongside the quantitative evaluation against the BS prototype (BS Proto.) in this supplement, demonstrates clear rendering and reconstruction advantages of isotropic over anisotropic Gaussian sampling. While a direct quantitative comparison with the official BS codebase on downstream tasks is precluded by practical incompatibilities, our proxy experiment confirms that anisotropic sampling significantly degrades reconstruction fidelity. We emphasize that our primary comparisons---against Berio~\etal~\cite{berioNeuralImageAbstraction2025} for calligraphy and against LIVE~\cite{maLayerwiseImageVectorization2022} for vectorization---are conducted under fair, controlled conditions with identical initialization and training iterations.

\end{document}